\documentclass[10pt]{article}
\usepackage[margin=1.0in]{geometry}
\usepackage{amsmath}
\usepackage{graphicx}
\usepackage{amsfonts}
\usepackage{amssymb}
\usepackage{bm}
\usepackage{hyperref}
\usepackage{natbib}
\usepackage[T1]{fontenc}

\usepackage{color}

\newcommand\T{\rule{0pt}{2.6ex}}       
\newcommand\B{\rule[-1.2ex]{0pt}{0pt}} 

\setcitestyle{authoryear}

\begin{document}

\title{\textbf{The Spectrum of the Universe}}

\author{Ryley Hill, Kiyoshi W. Masui and Douglas Scott \\
Dept.\ of Physics \& Astronomy, University of British
Columbia, Vancouver, BC, Canada}

\maketitle

\begin{abstract}
The cosmic background (CB) radiation, encompassing the sum of emission from
all sources outside our own Milky Way galaxy across the entire
electromagnetic spectrum, is a fundamental phenomenon in observational
cosmology.  Many experiments have been conceived to measure it (or its
constituents) since the extragalactic Universe was first discovered;
in addition to estimating the bulk (cosmic monopole) spectrum, directional
variations have also been detected over a wide range of wavelengths.
Here we gather the most recent of these measurements and discuss the current
status of our understanding of the CB from radio to $\gamma$-ray energies.
Using available data in the literature we piece
together the sky-averaged intensity spectrum, and discuss the emission
processes responsible for what is observed.  We examine the effect of
perturbations to the continuum spectrum from atomic and molecular line
processes and comment on the detectability of these signals.  
We also discuss how one could in principle obtain a complete census of the 
CB by measuring the full spectrum of each spherical harmonic expansion 
coefficient.
This set of spectra of multipole moments effectively encodes the entire
statistical history of nuclear, atomic and molecular processes in the
Universe.
\end{abstract}

\section{Introduction}
\label{introduction}

If you look at the sky with relatively wide angular resolution, concentrating
on radiation coming from beyond Earth's atmosphere, past the Solar System
and outside the Milky Way galaxy, you would only see diffuse anisotropies 
on top of a homogeneous, isotropic background. This is known as the ``cosmic 
background'' (CB). The sources contributing to this background can range from 
astronomically
small objects, such as atoms, nuclei and dust grains, to stars, galaxies 
and galaxy clusters, and are created by processes like nuclear fusion, 
gravitational collapse and thermal radiation. As the CB is composed of 
photons ranging from hundreds of thousands to nearly fourteen billion years 
old, its all-sky energy encodes the history of structure formation, energy 
distribution 
and expansion in the Universe. It is an important tool in cosmology today,
and a huge amount of effort has been put into measuring, interpreting and 
predicting its spectrum.

The CB can be observed over some 17 orders of magnitude in frequency 
($10^{8}$ to $10^{25}$\,Hz), which has required a plethora of detection 
techniques to measure and many different theoretical models to interpret,
hence we tackle this by splitting the spectrum into sections, or wavebands. 
Astronomers traditionally work in radio, microwave, infrared, optical, 
ultraviolet, X-ray and $\gamma$-ray wavebands, each section unique in its 
radiation detection methods, units and jargon, although the same physical 
quantity (namely the isotropic photon background) is being described. 
To construct the entire CB it will therefore be necessary to perform a 
survey across effectively the whole astronomical discipline and to understand 
exactly how each measurement has been made, while translating the results into 
something mutually comparable.

Historically \citep{Trimble2006}, the study of the CB took off in earnest with the discovery of
the diffuse microwave component in 1965 \citep{PW1965,DPRW1965}, although
the first cosmic sources of X-rays were known slightly earlier than that
\citep{Giacconi1962} and distant radio sources had been detected a couple of
decades earlier still, although their interpretation was a long and complicated story \citep[see e.g.][for a review]{Sullivan2009}, with direct detections of the radio background coming in the late 1960s \citep{Bridle1967}. Moreover, it had been known since the early 20th century that we could see
optical photons from well beyond our own Galaxy.  Emission from the Galaxy
is highly anisotropic, being concentrated into a band across the sky (the
Milky Way), which is visible in many wavebands.  The study of our own Galaxy,
and the objects within it, is of course an interesting topic in its own right,
but here we are focusing on more distant (`cosmic' or `cosmological')
radiation, and so we assume that the Galactic contributions have been removed.

Efforts to gather the necessary data into one coherent survey of the CB were 
first carried out by \citet{longair1972}. At that time, the shape of the 
background spectrum was largely speculative, with only a few rough
measurements available in selected wavebands, along with upper limits
at other wavelengths and theoretical ideas for what the backgrounds might be
from various physical processes.  Other studies tended to focus on one 
part of the electromagnetic spectrum, but about 20 years after Longair \& 
Sunyaev's seminal work, \citet{ressell1990} compiled an 
updated plot of the background intensity, showing the dramatic progress 
that had already been made.  Rather than
constructing their plot mostly from theoretical considerations, Ressell 
and Turner had enough physical data available to them that they could 
simply convert the units and put all the empirical information together.
Though a significant improvement on what had come before,
their work showed that there was still much uncertainty at key 
wavelengths. These results were later updated by \citet{scott2000}, and 
there have also been various compilations focusing on individual wavebands of 
the background, primarily in the optical/infrared 
\citep[e.g.,][]{longair2001,Hauser2001,Lagache2005,driver2016}, and the
X-ray/$\gamma$-ray
\citep[e.g.,][]{Boldt1987,Fabian1992,hasinger1996,bowyer2001,ajello2015b}
regimes, or carried out in the context of modelling studies
\citep[e.g.,][]{Bond1986,Gilli2007,Franceschini2008,Finke2010,Gilmore2012,khaire2018}.
These earlier CB studies either include measurements that are now quite 
out of date, or lack information from the wider range of frequencies available.
Some other recent reviews cover a range of wavelengths, but are less
comprehensive than this one
\citep[e.g.,][]{dwek2013,cooray2016,Franceschini2017}.

Because of the numerous applications of the CB in astronomy, astrophysics and 
cosmology, it is useful to compile an up-to-date picture of its 
intensity distribution. Here we combine the most recent published data
in order to present a single, coherent plot of the background 
radiation spectrum.  At the same time we explore some implications 
coming from picking apart the constituents of this spectrum.
We will begin in Section~\ref{unitsanddefinitions} with a brief 
discussion of what it means to measure the CB, what units are commonly 
adopted and various conversion factors between them.
In Section~\ref{thecosmicbackground} we present the sky-averaged CB
intensity in the different wavebands, and discuss some of the emission
processes responsible for the observed features. In
Section~\ref{lineemissionperturbations} we discuss 
perturbations to the CB contiuum spectrum from line emission processes, 
while in Section~\ref{beyondthemonopole} we consider a more complete picture 
of the background sky by considering higher-order multipole moments.
We summarize our results in Section~\ref{conclusions}. Our goal is to present
the current status of the spectrum of the 
Universe, along with a summary of our understanding of its origin and 
the implications it has for astrophysics and cosmology.

\section{Units and definitions}
\label{unitsanddefinitions}

It is first necessary to define {\it exactly what we mean by the CB}.
A measurement of background radiation sounds rather ambiguous. Where was 
the telescope pointed? What do we do about anisotropies? What about units? 
Larger detectors and bandwidths will receive more energy per second, larger 
beamwidths will see more of the sky simultaneously, so how do we interpret the 
numbers?

\subsection{Units}
\label{units}

We will report our findings in SI units of frequency $\nu$ in hertz and 
specific intensity $I_{\nu}$ in W\,m$^{-2}$\,sr$^{-1}$\,Hz$^{-1}$. 
This is the power a detector would receive per unit area, per solid angle 
observed on 
the sky, per unit frequency of light being detected. So, for example, 
multiplying $I_{\nu}$ by $4\pi$ (the solid angle of the entire sky) and 
by the area of a detector's surface and its frequency bandwidth would give the
amount of power that an ideal detector receives.

It is convenient to multiply $I_{\nu}$ by $\nu$ because $\nu I_{\nu}$ gives
the contribution per unit {\it logarithmic\/} scale. In addition, since
$I_{\nu}$ is proportional to the energy density per unit frequency, plotting
$\nu I_{\nu}$ will give a qualitative view of how the energy density is 
distributed across the spectrum. In other words, if two features have
equal width in $\log(\nu)$, then the one that is higher in $\nu I_\nu$
contains more energy density.

In the astronomical literature units vary wildly between waveband
specializations. 
Here we give some key conversion factors necessary to obtain the desired 
intensity unit, W\,m$^{-2}$\,sr$^{-1}$\,Hz$^{-1}$.

\subsubsection{Frequency}
\label{frequency}

Frequency, wavelength and energy are all interchangeable. The wavelength 
$\lambda$ is extensively used when discussing microwave through ultraviolet 
frequencies, but for radio waves the frequency tends to be used, and
when it comes to high-energy photons frequency is usually 
expressed as photon energy $E$. We convert wavelength to frequency via
\begin{equation}
\label{eq:lambdatonu}
\nu=\frac{c}{\lambda},
\end{equation}
where $c$ is the speed of light, and we convert energy to frequency with
\begin{equation}
\label{energytonu}
\nu = \frac{E}{h},
\end{equation}
where $h$ is Planck's constant. One also sometimes encounters the 
wavenumber, which is the frequency given in inverse length units
\begin{equation}
\label{wavenumbertonu}
{\tilde \nu} = \frac{1}{\lambda} = \frac{\nu}{c}.
\end{equation}

\subsubsection{Intensity}
\label{intensity}

If the specific intensity per unit {\it wavelength\/} is $I_{\lambda}$, 
we need to use
\begin{equation}
\label{IlambdatpInu}
I_{\nu} = \frac{\lambda^{2}}{c} I_{\lambda},
\end{equation}
to convert to specific intensity per unit frequency.

In radio astronomy, it is common to find the specific intensity given 
as a ``brightness temperature'' $T_{\rm b}$, which is defined as the
temperature of a perfect blackbody emitting at the observed intensity in the 
Rayleigh-Jeans (i.e., low energy) regime, i.e.,
\begin{equation}
\label{temperaturetoInu}
I_{\nu} \equiv \frac{2 \nu^{2} k T_{\rm b}}{c^{2}},
\end{equation}
where $k$ is the Boltzmann constant.

Low-energy astronomers also like to use the jansky (Jy) to represent specific 
flux (which is the specific intensity integrated over the solid angle of the 
region in the sky of interest), where
\begin{equation}
\label{janskytoInu}
1 \, \mathrm{Jy} \equiv 10^{-26} \, \mathrm{W \, m^{-2} \, Hz^{-1}}.
\end{equation}

\subsubsection{Energy}
\label{energy}

The erg is a commonly used quantity in several sub-branches of 
astronomy, while 
in high-energy astronomy energy is generally given in electron-volts 
(eV). In 
order to convert ergs to joules we use
\begin{equation}
\label{ergtojoule}
1 \, \mathrm{erg} \equiv 10^{-7} \, \mathrm{J},
\end{equation}
and in order to convert electron-volts to joules we use
\begin{equation}
\label{electronvolttojoule}
1 \, \mathrm{eV} \equiv (e/\mathrm{C}) \, \mathrm{J},
\end{equation}
where $e/\mathrm{C}$ is a unitless number whose magnitude is the electron
charge.

\subsubsection{Solid angle}
\label{solidangle}

For measuring radiation coming from the sky the {\it solid\/} angle is 
important.  This is often given in square degrees, $\mathrm{deg}^{2}$,
which can be converted to steradians through
\begin{equation}
\label{sqdegreetosteradian}
1 \, \mathrm{sr} \equiv \left( \frac{180}{\pi} \right)^{2} \, 
\mathrm{deg^{2}}.
\end{equation}

\subsection{Redshift}
\label{redshift}

In cosmology, distances are frequently referred to in terms of 
``redshift''. The redshift of a source is a measure of how much the observed
wavelengths are stretched by the expansion of the Universe:
\begin{equation}
\label{zdefinition}
1 + z = \frac{\nu_{\mathrm{em}}}{\nu_{\mathrm{obs}}}.
\end{equation}
Here $\nu_{\mathrm{em}}$ is the frequency of a photon that has 
been emitted by the source, and $\nu_{\mathrm{obs}}$ is the frequency that
same photon is observed to have on Earth. 

The evolutionary history of redshift, or equivalently expansion, is governed
by the now conventional vacuum-energy (or $\Lambda$) dominated cold dark matter 
($\Lambda$CDM) model, in which the Universe is flat on cosmological scales,
the energy density of the Universe is composed of (mostly cold dark) matter
and vacuum energy with fractions $\Omega_{\rm m} = 0.308 \pm 0.012$ and
$\Omega_{\Lambda} = 0.692 \pm 0.012$, respectively, and the local expansion
rate per unit length is
$H_{0} = (67.81\pm0.92)\,{\rm km}\,{\rm s}^{-1}\,{\rm Mpc}^{-1}$
\citep{planck2015cp}.
Because the expansion is so well understood, a measurement of the spectral
shift of a source tells us how distant the source is through a monotonic function:
\begin{equation}
\label{properdistance}
d(z) = \frac{c}{H_{0}} \int_{0}^{z} 
\frac{dz'}{(1+z')\sqrt{\Omega_{\rm m}(1+z')^{3}+\Omega_{\Lambda}}}.
\end{equation}
Redshifts are relatively easy to measure (from spectral lines) and then
distances are determined using the above equation.  As a result of the finite
speed of light, distant objects are observed as they were long ago.  Because
of this, cosmologists use redshift as a proxy for both distance and cosmic
epoch.

\subsection{Spherical harmonics}
\label{sphericalharmonic}

In general we could describe the sky by giving the spectral intensity in every
direction. The natural way to do this is to decompose the sky into spherical 
harmonics, i.e., the specific intensity as a function of polar angle $\theta$
and azimuthal angle $\phi$ can be written as 
\begin{equation}
\label{Inusphericalharmonic}
I_{\nu}(\theta,\phi) = \sum_{\ell=0}^{\infty} \sum_{m=-\ell}^{\ell} 
a_{\ell m} Y_{\ell}^{m}(\theta,\phi),
\end{equation}
where $Y_{\ell}^{m}(\theta,\phi)$ are the usual spherical harmonics.

Each term in the expansion has a physical meaning. Noting that $Y_{0}^{0}=1$, 
it follows that $a_{00}$ is just the intensity averaged over the sky.  The
three $\ell=1$ terms ($m=+1, 0, -1$) are related to the dipole moment of the 
intensity. For large $\ell$, the $\ell$th components of the 
spherical harmonic expansion correspond to the intensity contribution 
from angular scales of approximately $\pi/\ell$.

In what follows, we will be looking at the $\ell=0$ term, i.e.,
the average value of the intensity per unit frequency in the sky, 
after subtraction of all the light from the Milky Way, our Solar System and the 
atmosphere. This gives the average spectrum of radiation from the extragalactic 
Universe. We shall return to a discussion of angular dependence later -- but it 
is worth noting for now that since the Universe is close to homogeneous on 
large scales then any deviations from the average spectrum are relatively small.

\subsection{Dividing the CB}
\label{splittingthecb}

As we have previously mentioned, it is convenient to split the electromagnetic 
spectrum into sections. Following astronomical convention, these are the radio,
microwave, infrared, optical, ultraviolet, X-ray and $\gamma$-ray waveband
ranges; as such, we will denote the various sections of the CB
as the cosmic radio, microwave, infrared, optical, ultraviolet, X-ray and
$\gamma$-ray backgrounds, or the CRB, CMB, CIB, COB, CUB, CXB and CGB,
respectively. Table~\ref{cbsections}
lists these ranges in units of frequency, wavelength and energy.

\begin{table}[!t]
\begin{center}
\begin{tabular}{|l|c|c|c|}
\hline
Part of CB & Frequency range & Wavelength range  & Energy range \B\T \\
\hline
Cosmic radio background (CRB) & $<$ $10^{10}\,$Hz & $>30\,$mm & $<$ 40$\,\mu$eV  \T \\
\hline
Cosmic microwave background (CMB) & $10^{10}$--$10^{12}\,$Hz & 0.3--30$\,$mm & 
 0.04--4$\,$meV \T \\
\hline
Cosmic infrared background (CIB) & $10^{12}$--$10^{14}\,$Hz & 3--300$\,\mu$m & 
 4--400$\,$meV \T \\
\hline
Cosmic optical background (COB) & $10^{14}$--$10^{15}\,$Hz & 0.3--3$\,\mu$m & 
 0.4--4$\,$eV \T \\
\hline
Cosmic ultraviolet background (CUB) & $10^{15}$--$10^{16}\,$Hz & 30--300$\,$nm
 & 4--40$\,$eV \T \\
\hline
Cosmic X-ray background (CXB) & $10^{16}$--$10^{19}\,$Hz & 0.03--30$\,$nm & 
 0.04--40$\,$keV \T \\
\hline
Cosmic $\gamma$-ray background (CGB) & $>$ $10^{19}\,$Hz & $<$ 0.03$\,$nm & 
 $>40\,$keV \T \\
\hline
\end{tabular}
\caption[Cosmic background frequency sections]{Approximate divisions of the
cosmic backgound}
\label{cbsections}
\end{center}
\end{table}

\section{The cosmic background}
\label{thecosmicbackground}

We are now in a position to discuss the spectrum of the Universe. We 
will go over each division of the CB in some detail, discussing which physical 
processes are contributing to each frequency range and presenting the current
estimates of $\nu I_{\nu}$ from available data.  We will also describe
some of the measurement approaches
that are used to constrain the CB spectrum.  Most
astronomical imaging experiments are not sensitive to the average value of
the image; in other words they are effectively differential measurements,
which cannot directly determine the monopole amplitude.  There {\it are\/}
some absolute measurements of the amplitude of the CB, but these are the
exception, and, as we will see below, most determinations are more indirect.

\subsection[Cosmic radio background]{The cosmic radio background (CRB)\\
{\large [$\mathbf{\nu<10^{10}}\,$Hz, $\mathbf{\lambda>30}\,$mm}]}
\label{crb}

The radio portion of the spectrum includes all frequencies below around
$10^{10}\,$Hz. 
The current theoretical picture of the CRB is that it is mainly the sum 
of synchrotron emission emitted by charged particles moving through diffuse
galactic and inter-galactic magnetic fields, along with emission from
active galactic nuclei (AGN) and H\textsc{i} line emission, combined with the 
low energy end of the CMB \citep{singal2010}. The energy released by 
synchrotron radiation is proportional to the square of the magnetic field 
strength, but since magnetic fields on the scales of galaxies are quite small
\citep[of order $10^{-9}$\,T, e.g.,][]{widrow2002}, the photons emitted have
relatively low energies (see Section~\ref{cmb} for details on the CMB
contribution).

\begin{figure}[ht!]
\begin{center}
\includegraphics[scale=0.8]{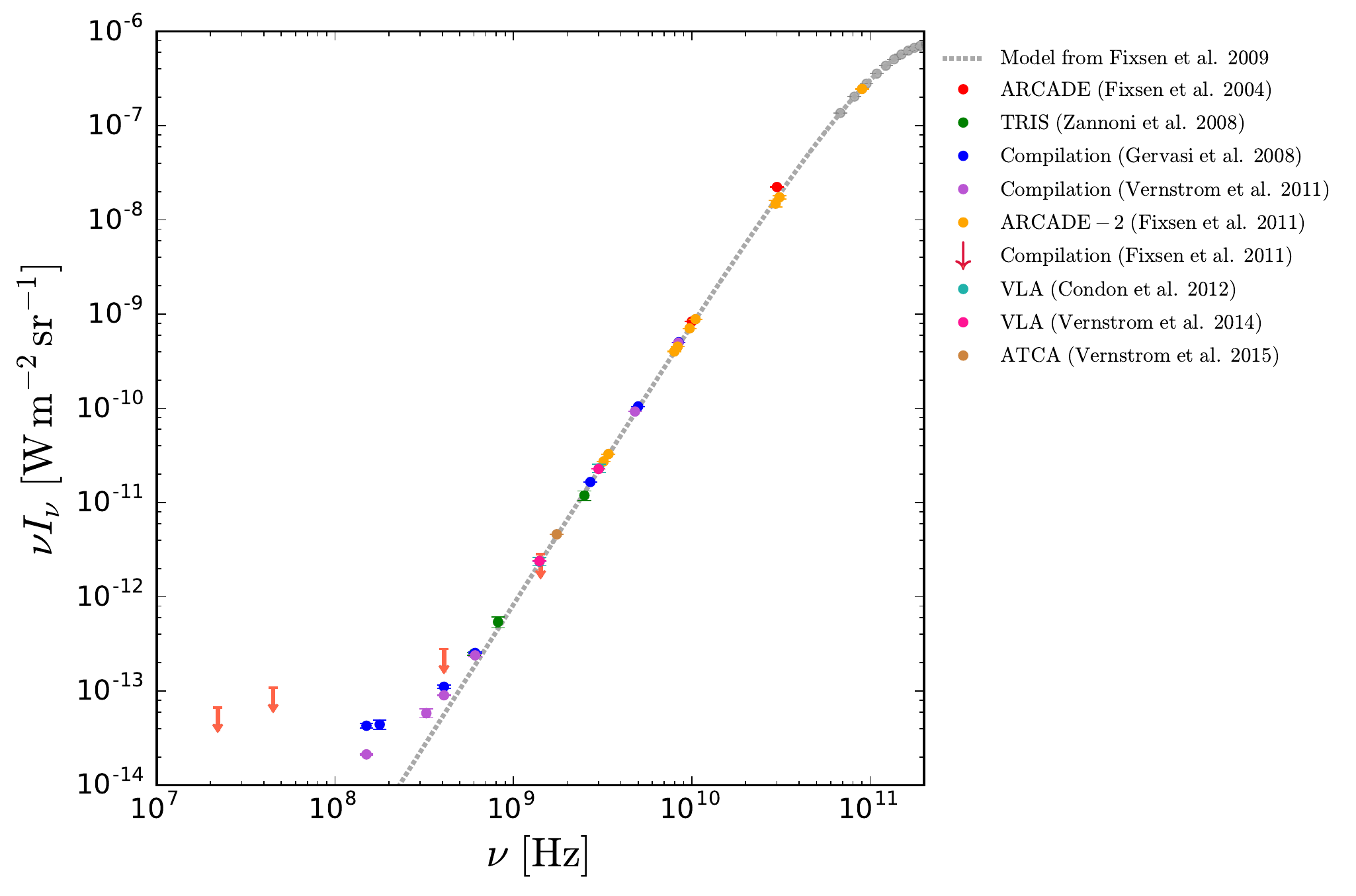} \caption[Cosmic radio background]
{\label{crbfig}
Cosmic radio background radiation. This is the low energy end of the CMB, with 
a slight excess from various astrophysical phenomena such as synchrotron 
emission, AGN and H\textsc{i} emission. A 2.7255\,K blackbody curve is shown
as the dotted line for comparison. The downward pointing arrows represent upper
limits on the background from measurements of the absolute intensity of the sky
with modelled foregrounds subtracted. The faded grey points correspond to
values outside the radio frequency range.}
\end{center}
\end{figure}

There is a lower limit to the frequencies of the CRB that we can observe
due to the plasma frequency of the Earth's atmosphere and our Milky Way Galaxy, given by
\begin{equation}
\label{plasmafrequency}
\nu_{\rm p} = \sqrt{\frac{n_{\rm e} e^{2}}{\pi m_{\rm e}}},
\end{equation}
where $n_{\rm e}$ is the electron number density and $m_{\rm e}$ the electron 
mass. This frequency is the rate at which free electrons in a plasma 
oscillate; if the frequency of a propagating electromagnetic wave is less than 
the plasma frequency, it will be exponentially attenuated on scales much 
shorter than the wavelength \citep{chabert2011}. The Earth's 
ionosphere has a plasma frequency on the order of 10\,MHz, limiting all ground 
based observations of the CRB, while the interstellar medium (ISM) 
has a plasma frequency of about 1\,MHz, limiting even space-based observations 
\citep{lacki2010}.

The data in this waveband come from a combination of ground and balloon-borne 
instruments. In our compilation (see Fig.~\ref{crbfig})
we include measurements of
the absolute intensity of the sky from the Absolute Radiometer for 
Cosmology, Astrophysics and Diffuse Emission (ARCADE) experiment 
\citep{fixsen2004} and its successor, ARCADE-2 \citep{fixsen2011},
which measured the CRB between 3 and 90\,GHz by subtracting a model of Galactic
emission. Also reported 
by these authors are four absolute measurements performed between 1980 and 1990
at 0.022, 0.045, 0.408 and 1.42\,GHz, to which the same Galactic foreground
model was subtracted to obtain the CRB\@. We consider the latter four measurements as {\it upper\/} limits because these observations were taken much closer to the Galactic plane than later dedicated background experiments.

In addition, the TRIS experiment, a set
of three ground based radiometers, measured the CRB at 0.6, 0.8 and 2.5\,GHz 
\citep{zannoni2008}. We also use two studies that employed a compilation of
radio galaxy counts to estimate the total value of the CRB
\citep[][and references therein]{gervasi2008,vernstrom2011}. We give further
derived detections of the CRB obtained with the Karl G\@. Jansky Very Large Array
(VLA) at 1.4 and 3\,GHz 
\citep{condon2012,vernstrom2014} and by the Australian Telescope Compact
\citep[ATCA,][]{vernstrom2015} at 1.75\,GHz.

The values we report here were converted from
brightness temperature (see Eq.~\ref{temperaturetoInu}). We note 
that the CRB value reported by \citet{vernstrom2011} includes {\it only\/} the 
contribution from galaxies (and would have missed any genuinely diffuse
emission). It is therefore necessary to add the CMB temperature, 
experimentally found to be $T_{\mathrm{CMB}}=2.7255\pm0.0006$\,K
\citep[][as discussed in the next sub-section]{fixsen2009}, to obtain the
total intensity. 

Figure~\ref{crbfig} shows the CRB (i.e., the CB at radio frequencies).
For comparison, we have plotted the low-energy end of a blackbody spectrum
with temperature $T_{\mathrm{CMB}}$.  We can see very good 
agreement in the shape of the curve with that of a blackbody over much of the
radio region, and as expected, at lower frequencies the data lie slightly
higher due to the contributions from galaxies. There is some debate regarding
the robustness of the excess radiation seen by the ARCADE-2 experiment
\citep[e.g.,][]{subrahmanyan2013}, in particular at the higher
frequencies. Subtracting $T_{\mathrm{CMB}}$ from the brightness temperatures 
obtained from each experiment shows that ARCADE-2 obtained excess values that 
were significantly higher than those observed by TRIS or those calculated via
galaxy counts (see \citealt{vernstrom2011} and \citealt{singal2017} for more
detailed discussions).

\subsection[Cosmic microwave background]{The cosmic microwave background
 (CMB)\\ 
{\large [$\mathbf{\nu=10^{10}}$--$\mathbf{10^{12}}\,$Hz,
 $\mathbf{\lambda=0.3}$--30\,mm}]}
\label{cmb}

The CMB is dramatically the highest amplitude part of the CB\@.  It is also
the most thoroughly studied portion of the spectrum (possibly the most
thoroughly studied phenomenon in cosmology), and its origin is now very 
well understood. The CMB is the blackbody radiation left over from the hot 
early phase of the Universe. It was last scattered by matter about 400{,}000 
years after the Big Bang, when the temperature cooled enough to allow protons 
to combine with electrons and form neutral atoms, transforming the Universe 
from optically thick to optically thin. The temperature of the Universe at this 
time was about 3000\,K\@. Today we see the temperature to be 2.7255\,K,
owing to the fact that scales have redshifted by a factor of about 1100
since the last-scattering 
epoch \citep[see e.g.,][]{ScottSmoot2016}.  A great deal of physics can be
discerned from studying the possible spectral distortions and spatial
variations (or anisotropies) in the CMB \citep[e.g.,][]{samtleben2007}.
These anisotropies currently provide the tightest constraints on the
parameters that statistically describe our Universe \citep{planck2015cp}.

\begin{figure}[ht!]
\begin{center}
\includegraphics[scale=0.8]{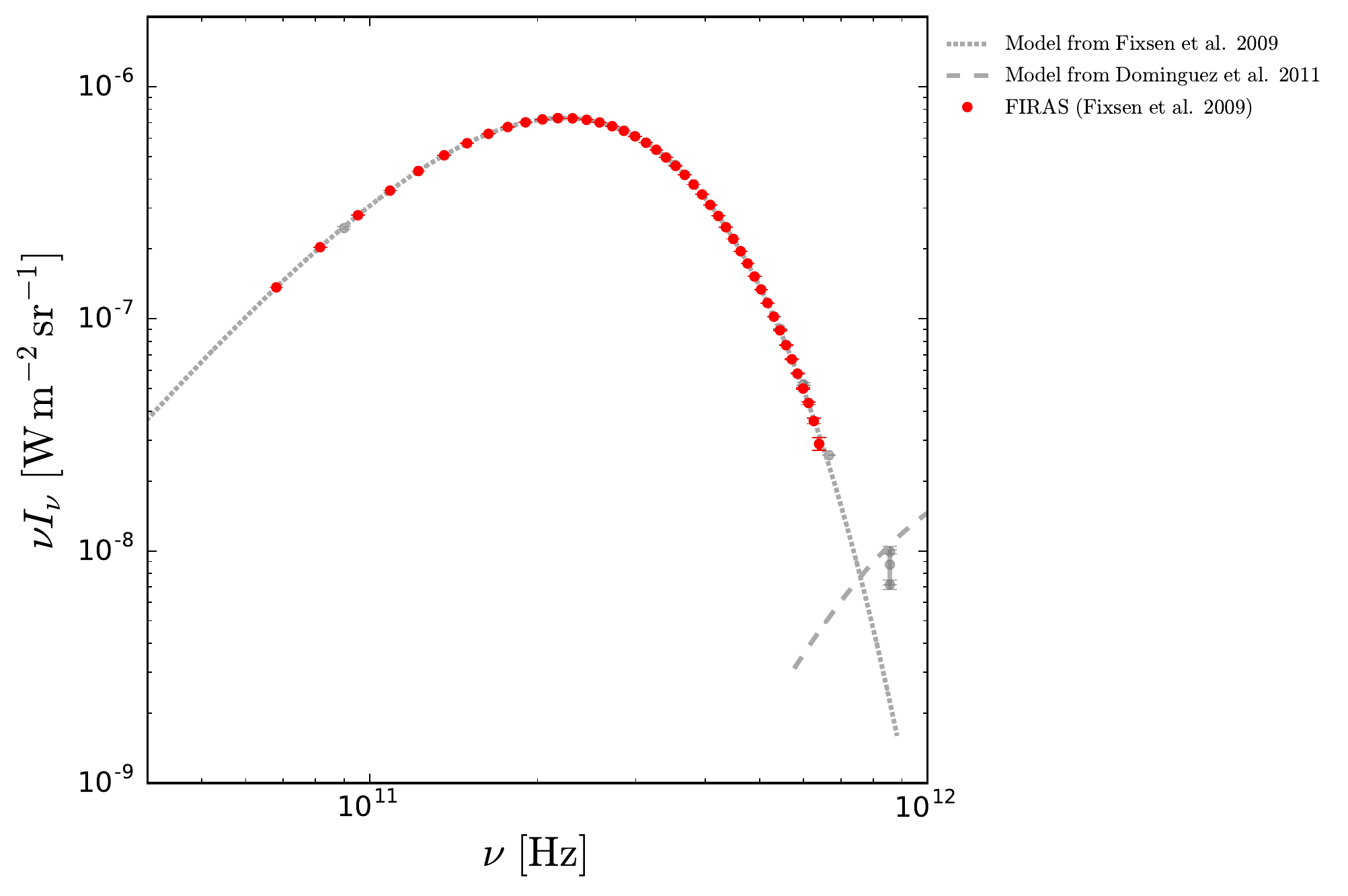} \caption[Cosmic microwave 
background]
{\label{cmbfig}
Cosmic microwave background radiation. A blackbody spectrum at 2.7255\,K is 
plotted as the dotted line to demonstrate the agreement between theory 
and observation in this part of the spectrum. The faded grey points correspond
to values outside the microwave frequency range.}
\label{cmbfig}
\end{center}
\end{figure}

Blackbody radiation from the CMB follows the Planck intensity function,
\begin{equation}
\label{blackbody}
I_{\nu} = \frac{2 h \nu^{3}}{c^{2}} \frac{1}{e^{h \nu / k_{\mathrm{B}} T} -1}.
\end{equation}
For a blackbody with $T_{\mathrm{CMB}}=2.7255$\,K, the peak (in $\nu I_\nu$)
is at about $220\,$GHz, or $1.4\,$mm.  Given the precise mathematical form
of this part of the CB, we can integrate to obtain the energy density of the
CMB, $u_{\rm CMB}=4.2\times10^{-14}$\,J\,m$^{-3}=
0.26$\,eV\,cm$^{-3}$, or the number density of CMB
photons, $n_{\rm CMB}=410$\,cm$^{-3}$.

The CMB spectrum has been measured to great precision by the Far InfraRed 
Absolute Spectrophotometer (FIRAS) instrument aboard the {\it COsmic Background 
Explorer\/} ({\it COBE\/}) satellite \citep{fixsen1996},
with a spectrum shown to be
very close to that of a blackbody. There are also earlier measurements that
have provided a useful supplement to
those from {\it COBE}-FIRAS, particularly at lower frequencies
\citep[e.g.,][as well as those in the radio, as already
discussed]{halpern1988,gush1990,mather1990,levin1992,staggs1996}; however,
we have not incorporated these into our analysis, since the measurements from
FIRAS span the same frequency range and provide smaller error bars.
On the high frequency side,
additional measurements at high resolution provide estimates of the resolved
(i.e., broken up into discrete sources)
fraction of the background at sub-millimetre wavelengths, and in particular,
recent observations using the Atacama Large Millimeter/sub-millimeter Array
(ALMA) have resolved close to 100$\%$ of the CMB-subtracted background
into extragalactic sources
\citep{carniani2015,fujimoto2016,aravena2016}. However, due to the fact that the
contribution to the CMB from extragalactic sources is several orders of
magnitude smaller than the thermal radiation leftover from the Big Bang,
we have not incorporated these measurements into our compilation.

The CMB measurements from FIRAS are plotted in Fig.~\ref{cmbfig},
along with a blackbody function 
(Eq.~\ref{blackbody}) with a temperature of $T_{\mathrm{CMB}}$.  Error
bars are present, but in many cases are much smaller than the points
themselves.  Agreement between the measurements and a Planck spectrum
is one of the strongest pieces of evidence for the hot Big Bang model, since
the photon spectrum is expected to naturally relax to a state of thermal
equilibrium in a model in which the early Universe was hot and dense; there is
no other explanation for this nearly perfect blackbody shape.

\subsection[Cosmic infrared background]{The cosmic infrared background (CIB)\\ 
{\large [$\mathbf{\nu=10^{12}}$--$\mathbf{10^{14}}\,$Hz,
 $\mathbf{\lambda=3}$--300\,$\mathbf{\mu}$m}]}
\label{cib}

The CIB contains approximately half of the total energy density of the
radiation emitted by 
stars through the history of the Universe (although with roughly a factor of 40
less total energy density than contained in the CMB)
and is tightly linked to the history 
of galaxy formation. This radiation is primarily emitted by dust that has been 
heated by the stars contained within galaxies \citep[e.g.,][]{Lagache2005}.
At the long wavelength end we see the energy output of the oldest, most 
redshifted galaxies, which are key to understanding the beginnings of galaxy
evolution.

\begin{figure}[ht!]
\begin{center}
\includegraphics[scale=0.8]{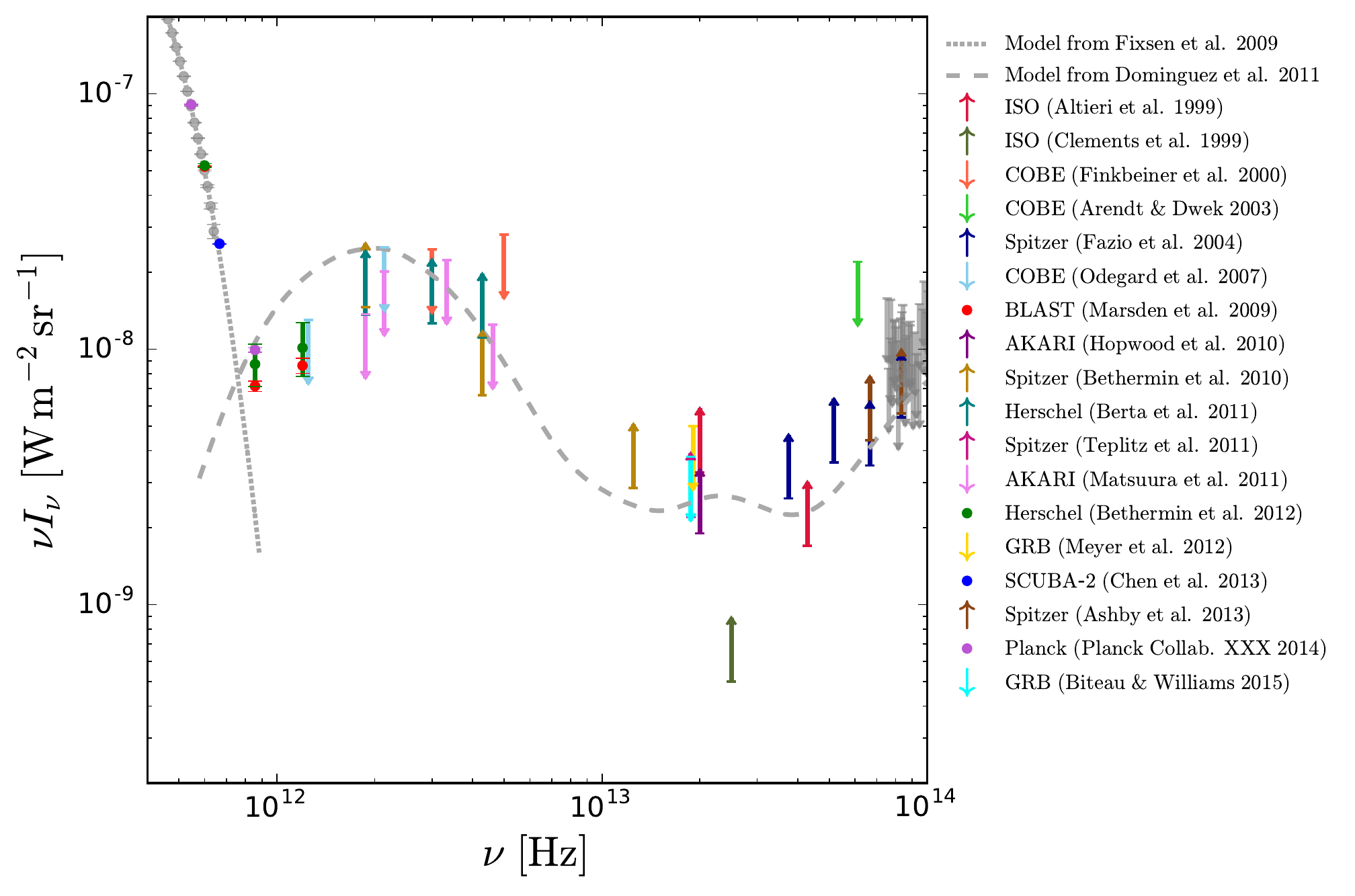} \caption[Cosmic infrared background]
{\label{cibfig}
Cosmic infrared background radiation. This region contains the second peak
in the CB, which arises from emission by dust re-radiating stellar emission 
in galaxies. A blackbody spectrum at 
2.7255\,K is plotted as a dotted line, and the dashed line is an
example of a model of the CIB/COB, derived by making a weighted sum of
the observed spectra of galaxies \citep{Dominguez2011}. The downward pointing
arrows represent upper limits on the background from measurements of the
absolute intensity of the sky with modelled foregrounds subtracted, while the
upward pointing arrows represent lower limits on the background from adding
contributions from resolved sources.}
\label{cibfig}
\end{center}
\end{figure}

This portion of the CB is extremely difficult to measure through absolute 
intensity measurements. Objects in our Solar System (such as planets, asteroids 
and dust) contribute a huge amount of unwanted `zodiacal' emission, and dust emission from our
own Milky Way Galaxy provides orders of magnitude of foreground emission to 
subtract from the measurements \citep[e.g.,][]{whittet1992}. One approach to deal with
this difficulty is to count the number of resolvable galaxies and measure their 
flux, extrapolate the data to take into account fainter undetected
sources, then sum 
over the whole sky \citep[as was done by][with radio galaxies]{vernstrom2011}. 
This gives a {\it lower\/} estimate, since it deals with only the contributions 
to the CIB from individually identified sources. {\it Upper\/} bounds,
on the other hand, can be obtained through an entirely different approach.
The spectra of $\gamma$-ray-emitting objects (such as a particular class of AGN
called ``blazars'') should have a deficit of high-energy photons, which
interact with low energy CIB photons via pair production \citep[see][for 
details]{dejager1992}. However, such upper limits depend of how the intrinsic
spectra of blazars are modelled, as well as details of the distribution of
sources responsible for producing the CIB photons, and hence
there is not complete agreement on how tight the limits are.  With these
complications in mind we will report only a few representative
results on upper or lower bounds to the CIB\@.

In the far- to mid-IR range (comprising 500--100\,$\mu$m) we use measurements 
taken by adding up sources detected by
the {\it Herschel Space Observatory}'s Photodetector Array Camera and 
Spectrometer \citep[PACS,][]{berta2011} and Spectral and Photometric 
Imaging REceiver \citep[SPIRE,][]{bethermin2012}, and the Balloon-borne Large
Aperture Submillimeter Telescope \citep[BLAST,][]{marsden2009}.  There have
also been estimates of the CIB at 450 and 850\,$\mu$m made using the
Submillimeter Common-User Bolometer Array-2 (SCUBA-2), a camera mounted
on the James Clerk Maxwell Telescope \citep[JCMT,][]{chen2013}. In addition,
we include two absolute measurements of the CIB at 140 and 240\,$\mu$m, made
with the Diffuse InfraRed Background Experiment (DIRBE) aboard {\it COBE\/}
\citep{odegard2007}. We also have estimates using galaxy counts from the 
Far-Infrared Surveyor (FIS) aboard the {\it Akari\/} satellite at
65, 90, 140 and 160\,$\mu$m \citep{matsuura2011}, and more recently from
the {\it Planck\/} satellite between 350 and 1400\,$\mu$m \citep{planck2014a}.
The CIB values determined in 
most of these papers had the CMB subtracted, since the scientific motivation 
was to study excess intensity over the CMB blackbody as opposed to the total 
intensity; for our purposes we have added a 
blackbody with $T_{\mathrm{CMB}}$ back into the results.

Moving on to the mid-infrared (100--5\,$\mu$m), further analyses of 
DIRBE data at 4.9\,$\mu$m, and at 60 and 100\,$\mu$m by \citet{arendt2003} and 
\citet{finkbeiner2000}, respectively, have yielded more robust background
estimates than the earlier studies.  Additionally, the {\it Spitzer Space
Telescope\/} was used to observe the infrared sky at 3.6, 4.5, 5.8 and
8.0\,$\mu$m with the InfraRed Array Camera (IRAC), from which the CIB
intensity was estimated by summing up detected sources
\citep{fazio2004,ashby2013}. The Multiband Imaging Photometer for Spitzer
(MIPS), with wavebands centred at 24, 70 and 160\,$\mu$m, was used in a similar
fashion to obtain additional CIB measurements \citep{bethermin2010}.
{\it Spitzer}'s final instrument, the InfraRed Spectrograph (IRS), was
employed by \citet{teplitz2011} to estimate the CIB at 16\,$\mu$m.
To fill in the gap between IRAC and MIPS, we present further CIB 
lower estimates from the {\it Infrared Space Observatory\/} ISOCAM at 7, 12 and 
15\,$\mu$m \citep{altieri1999,clements1999}, and {\it Akari}'s
InfraRed Camera \citep[IRC,][]{hopwood2010}. 

Upper bounds found from $\gamma$-ray spectra of blazars have been derived using 
many different observations and detailed modelling approaches.
Here we are primarily interested in 
obtaining upper limits in the mid-IR range, where dust contamination prevents 
absolute measurements from being obtained. For this, \citet{meyer2012} derived 
an upper limit of 24\,$\mathrm{nW \, m^{-2} \, sr^{-1}}$ at 8\,$\mu$m, while 
\citet{biteau2015} obtained a limit of 16\,$\mathrm{nW \, m^{-2} \, sr^{-1}}$ 
at 16\,$\mu$m. There are many other $\gamma$-ray-derived background limits that 
could be used \citep[e.g.,][]{dwek2013}, but for the sake of clarity we will
only include the two above.

The total CIB is shown in Fig.~\ref{cibfig}, with the CMB blackbody shown again
for comparison. There have been many attempts to build models for the
infrared/optical background. We have plotted one example of such models, which
was constructed by adding up the spectral energy distributions of a set of
galaxy types, weighted by appropriate fractions \citep{Dominguez2011}.
There is a clear local maximum here, despite the 
constraints only coming from upper and lower bounds in some regions.
The peak lies at around 2--$3\times 10^{12}$\,Hz (100--150\,$\mu$m) and
contains an energy density of approximately $10^{-15}\,{\rm J}\,{\rm m}^{-3}$.

\subsection[Cosmic optical background]{The cosmic optical background (COB)\\ 
{\large [$\mathbf{\nu=10^{14}}$--$\mathbf{10^{15}}\,$Hz,
 $\mathbf{\lambda=0.3}$--3\,$\mathbf{\mu}$m}]}
\label{cob}

The optical background is perhaps the second most important
portion of the CB (at least for beings with eyes like ours) after the CMB,
since it is dominated by the emission coming 
directly from stars. From this radiation we can infer much about the history of 
cosmic star formation.  Because of this, it is a highly
studied region of the CB -- in addition to the fact that astronomy has
traditionally been dominated by optical wavelengths, driven by the fact that
ground-based telescopes can easily study optical photons.

\begin{figure}[ht!]
\begin{center}
\includegraphics[scale=0.8]{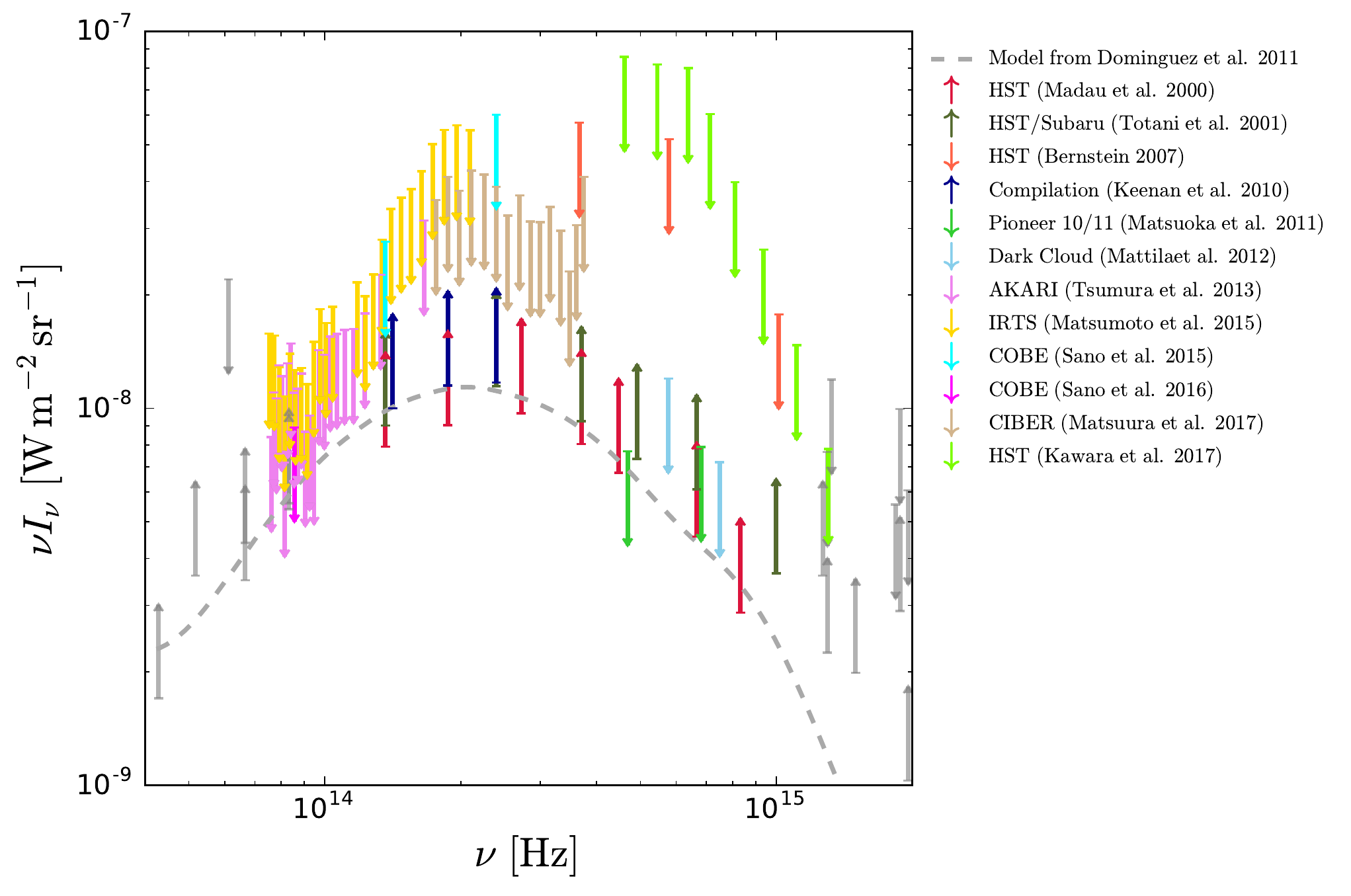} \caption[Cosmic optical background]
{\label{cobfig}
Cosmic optical background radiation. This region contains the third 
peak in the CB, at around $3\times10^{14}$\,Hz (1\,$\mu$m), corresponding to 
emission originating from nucleosynthesis in stars.  The dashed line is an
example of a model of the CIB/COB, derived by making a weighted sum of
the observed spectra of galaxies \citep{Dominguez2011}.}
\end{center}
\end{figure}

Since the Milky Way's dust emission continues to be a confusing foreground well
into optical wavelengths, direct estimates of the COB are also challenging.
The best measurements tend therefore to be sums over sources. For instance, 
\citet{madau2000} achieved lower bounds on the COB with the {\it Hubble Space 
Telescope}'s ({\it HST\/}) Hubble Deep Field survey in seven
optical bandpasses (0.36--2.2\,$\mu$m) by counting galaxies.  In 
addition, \citet{totani2001} included in the galaxy counts sources observed 
in the Subaru Deep Field by the ground-based Subaru Telescope. An attempt to
obtain absolute background estimates using {\it HST\/}
observations between 0.3 and 0.8\,$\mu$m was carried out by 
\citet{bernstein2007}, and more recently between 0.2 and 0.7\,$\mu$m
by \citet{kawara2017}, which we treat here as upper limits. Lastly, 
\citet{keenan2010} calculated the COB contribution from galaxies using galaxy 
counts from a wide range of studies dating back to 1999 (including their own 
observations in the near-IR $K$ band with Subaru).

There have been many additional attempts to determine the COB spectrum via 
direct detection, i.e., by attempting to subtract Galactic and zodiacal 
emission from rocket or satellite data.
We have taken data obtained by the IRC mounted aboard the {\it Akari\/}
satellite \citep{tsumura2013} and by the Near Infra-Red Spectrometer (NIRS)
on the {\it InfraRed Telescope in Space\/} \citep[IRTS,][]{matsumoto2015}, each
of which was designed to study 1.5--4.0\,$\mu$m radiation. More recently than
these experiments was the Cosmic Infrared Background Explorer (CIBER),
primarily designed to 
measure {\it fluctuations\/} around 1\,$\mu$m, but which provided an 
estimate of the averaged power using models of the fractional level of 
fluctuations \citep{matsuura2017}. There has also been renewed interest in 
eliminating systematic effects from the DIRBE observations;
re-analyses were performed at 
1.25, 2.2 and 3.5\,$\mu$m \citep{sano2015,sano2016}. A further approach to 
remove zodiacal contamination involved using background light measurements from 
Pioneer 10 and 11 Imaging Photopolarimeter (IPP) data \citep{matsuoka2011}; the 
instruments observed the sky when the spacecraft were about 5\,AU from the Sun, 
where zodiacal light contributions were smaller. Lastly, we incorporate a
measurement at 400 and 520\,nm using the ``dark cloud'' 
technique \citep{mattila2012}, where 
the difference in intensity between a high latitude dark nebula and its 
surroundings is taken to be solely due to the COB\@.

These results are combined in Fig.~\ref{cobfig}. Following the trough seen in
the CIB, there is another peak here at about $3\times10^{14}$\,Hz (1\,$\mu$m),
containing a similar total energy density to that from the CIB peak
(i.e., around $10^{-15}\,{\rm J}\,{\rm m}^{-3}$).
There is some tension between the lower limits from galaxy counts, the 
``dark cloud'' measurement and the analysis of Pioneer 10 and 11 data. 
Systematic effects are difficult to control in direct measurement techniques, 
and it seems likely that the estimates from galaxy counts are close to the true 
background levels. We also show in Fig.~\ref{cobfig} the model from
\citet{Dominguez2011}, which carries over to optical frequencies. In 2002 a
similar study, adding up the spectra of 200{,}000 galaxies, suggested
that the average optical colour of the Universe is turquoise; however, this
was quickly corrected (due to a coding error) to a slightly off-white colour,
corresponding to what one might call `beige' \citep{Baldry2002}.  Other
attempts at modelling the optical and IR backgrounds have made use of
evolutionary models constrained by existing data or integration over galaxy
luminosity functions, which find broadly similar curves to the
one plotted in Figs.~\ref{cibfig} and~\ref{cobfig}
\citep[e.g.,][]{Franceschini2008,Inoue2013,Stecker2016}.

\subsection[Cosmic ultraviolet background]{The cosmic ultraviolet 
 background (CUB)\\ 
{\large [$\mathbf{\nu=10^{15}}$--$\mathbf{10^{16}}\,$Hz,
 $\mathbf{\lambda=30}$--300\,nm}]}
\label{cub}

The next frequency range corresponds to the CUB, encompassing 
$10^{15}$--$10^{16}$\,Hz. The origin of this background is largely the light of 
hot, young stars and interstellar nebulae, including scattering by dust
(as opposed to absorption and re-emission), with contributions from hot
inter-cluster gas \citep[but the level of this
is still under some debate, see e.g.,][]{henry2015}. It should not come 
as a surprise that the CUB remains the most poorly studied portion of the CB to 
date because of neutral hydrogen's efficiency at absorbing UV light 
\citep[thus rendering the ISM nearly opaque at these frequencies,
e.g.,][]{bowyer2000}, and because of 
the fact that one must leave the Earth's atmosphere to study these wavelengths.
As a result, the available data contains levels of systematic uncertainty
that are hard to quantify, and it seems best to treat measurements as
upper (or sometimes lower) limits on the CUB\@.  The upper limits
come from photon flux counts carried out between 1970 and 1990, which were 
actually measuring the {\it total\/} UV background, including Galactic and 
zodiacal contributions, while the few available lower limits come from the 
usual galaxy counts (which do not go particularly faint).
There is plenty of room for improvement here. 

\begin{figure}[ht!]
\begin{center}
\includegraphics[scale=0.8]{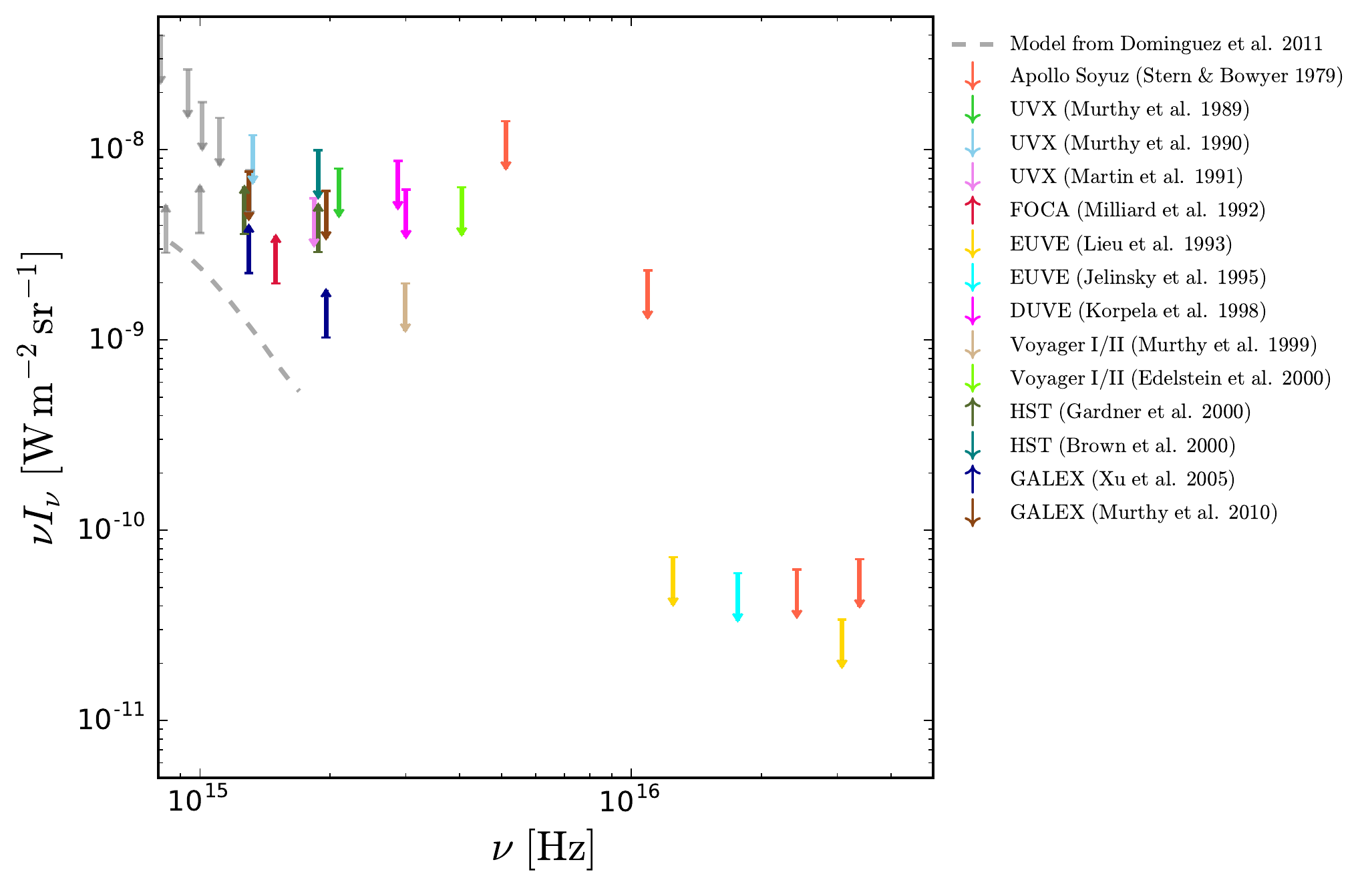} \caption[Cosmic ultraviolet 
background]
{\label{cubfig}
Cosmic ultraviolet background radiation. The sources of emission responsible
for this portion of the spectrum are primarily hot young stars and hot
inter-cluster gas. The downward pointing arrows represent upper limits on the
background from measurements of the absolute intensity of the sky with modelled
foregrounds subtracted, and the upward pointing arrows represent lower limits
on the background from adding contributions from resolved sources.}
\end{center}
\end{figure}

At the low end of the frequency range there have been several estimates of the
CUB derived from galaxy counts using the Space Telescope Imaging Spectrograph
(STIS) on {\it HST\/}
\citep{gardner2000,brown2000} and the {\it GALaxy Evolution eXplorer} satellite
\citep[{\it GALEX},][]{xu2005}, which we give as lower limits to the total CUB,
since these estimates lack the diffuse contribution from inter-cluster gas.
We also include an earlier 
estimate using a balloon-borne UV-imaging telescope centred at 200\,nm 
\citep{milliard1992}. For upper limits where Galactic emission must be
subtracted, {\it GALEX} has also provided the most recent estimate of the CUB
\citep{murthy2010}. Prior to this, we include contributions from the following
experiments: the Ultraviolet Spectrometer (UVS) aboard 
the {\it Voyager\/} spacecraft \citep{murthy1999,edelstein2000};
the Diffuse Ultraviolet Experiment \citep[DUVE,][]{korpela1998};
and a series of UV instruments flown
aboard the Space Shuttle {\it Columbia\/} in 1986 
\citep{murthy1989,murthy1990,martin1991}.  At the high frequency end of
the range (tens of nanometres), estimates have been taken from measurements
made on the {\it Apollo- Soyuz\/} mission by the Extreme UltraViolet Telescope
\citep[EUVT,][]{stern1979}, covering 9--59\,nm. 
Lastly we report findings on the total diffuse UV background (so again, upper 
limits on the CUB) from the Extreme UltraViolet Explorer's (EUVE) Deep Survey 
(DS) telescope at 10, 17 and 24\,nm \citep{lieu1993,jelinsky1995}.

Figure~\ref{cubfig} displays the current status on the CUB, showing a
general decline in intensity from beginning to end, with no interesting 
features visible due to the lack of constraining data at these 
wavelengths.

\subsection[Cosmic X-ray background]{The cosmic X-ray background (CXB)\\
{\large [$\mathbf{\nu=10^{16}}$--$\mathbf{10^{19}}\,$Hz,
 $\mathbf{\lambda=0.03}$--30\,nm}]}
\label{cxb}

As we move to even higher energies we arrive at the CXB, spanning the range 
from $10^{16}$ to $10^{19}$\,Hz. It is now generally believed that the
astrophysical processes producing the CXB are dominated by the accretion disks
around AGN \citep[e.g.,][]{comastri1995}, which are hot enough to emit thermal
bremsstrahlung photons observed in the X-ray range.
This source of radiation is negligible at lower frequencies, where
we largely see thermal energy (leftover from the early Universe and from warm
dust) and nuclear energy (released in the form of photons from nuclear
reactions in the cores of stars); however, at X-ray and $\gamma$-ray
frequencies gravitational energy is the dominant mechanism sourcing the
spectrum through accretion of gas.

\begin{figure}[ht!]
\begin{center}
\includegraphics[scale=0.8]{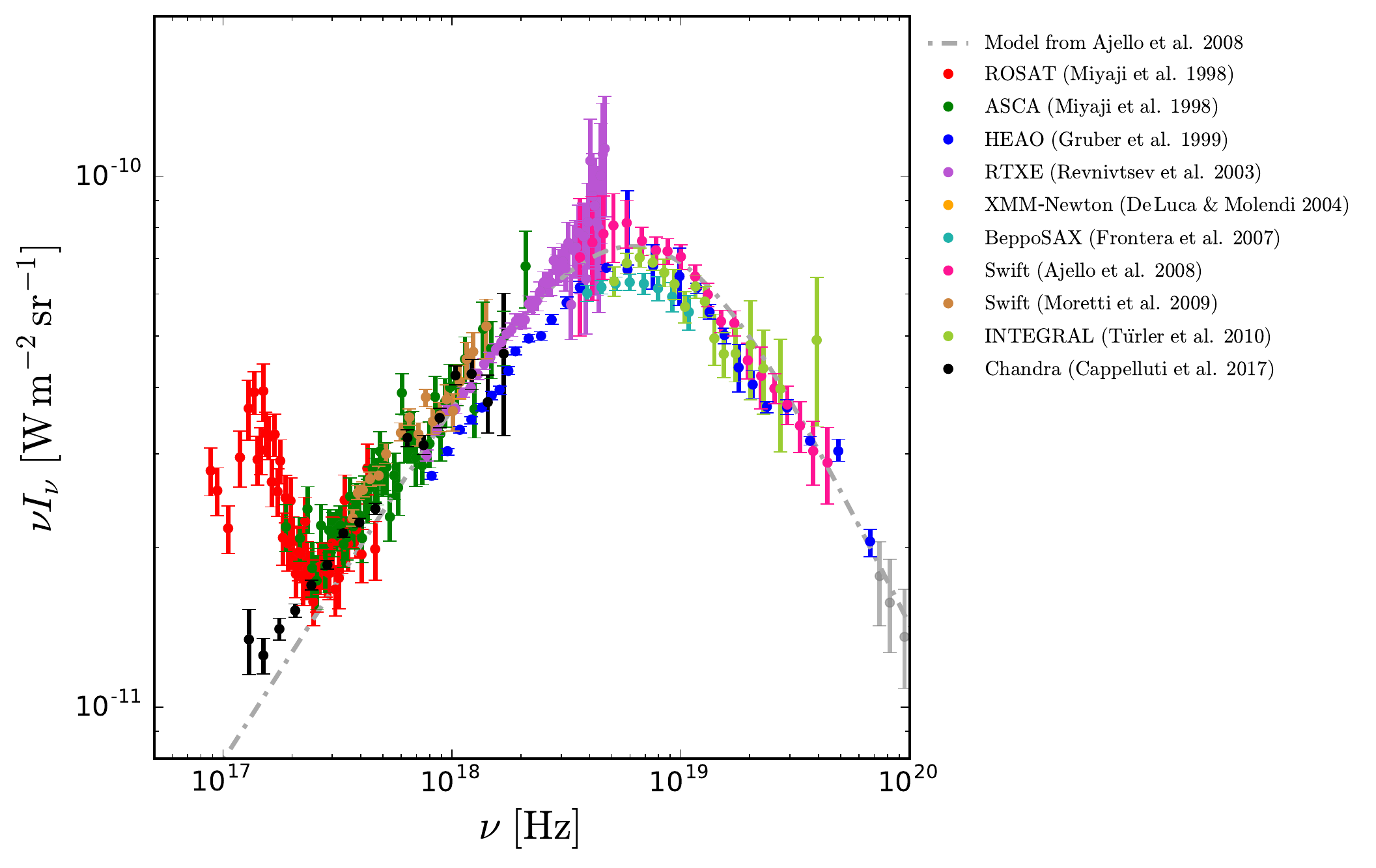} \caption[Cosmic X-ray background]
{\label{cxbfig}
Cosmic X-ray background radiation. There is a maximum located at about $5 
\times 10^{18}$\,Hz (20\,keV). This part of the CB is dominated by 
gravitational energy that has been converted to radiation by the ultra-hot 
accretion disks around AGN\@.  It is fairly well fit by a simple thermal
bremsstrahlung spectrum; the dot-dashed line shows a similar double
power-law model from \citet{ajello2008}.}
\end{center}
\end{figure}

There have been lots of measurements of the CXB within the past couple of 
decades, and it contains yet another peak in the CB spectrum, with important
astronomical and cosmological implications. Between 1 and 10\,keV
(electron-volts being the primary unit used by X-ray 
astronomers) we will use a joint data analysis from \citet{miyaji1998} that 
combined observations form two Proportion Sensitive Proportional Counters 
(PSPC) on {\it ROSAT\/} and two Gas Imaging Spectrometers (GIS) on the
{\it Advanced Satellite for Cosmology and Astrophysics\/} ({\it ASCA\/}),
plus data from {\it Swift}'s X-Ray 
Telescope \citep[XRT,][]{moretti2009}, and {\it XMM-Newton}'s European
Photon Imaging Camera \citep[EPIC,][]{deluca2004}. There also exist CXB
observations from {\it Chandra}'s Advanced CCD Imaging Spectrometer (ACIS),
described by \citet{hickox2006} and later updated by \citet{cappelluti2017};
we use the latter work here (although they are largely consistent with each
other).  Next, between 3 and 20\,keV, we 
include the \citet{revnivtsev2003} CXB results obtained from the Rossi X-ray 
Timing Explorer (RXTE) using the Proportion Counter Array (PCA), and 
between 16 and 45\,keV results from the {\it BeppoSAX\/}
Phoswich Detection System \citep[PDS,][]{frontera2007}. 

Moving into the hard X-ray portion of the spectrum, the CXB has been observed 
from 15 to 180\,keV by the {\it INTErnational Gamma-Ray Astrophysics
Laboratory\/} ({\it INTEGRAL\/}) via the Imager on-Board the INTEGRAL Satellite
\citep[IBIS,][]{turler2010}, and another instrument of {\it Swift}'s,
the Burst Array Telescope \citep[BAT,][]{ajello2008}.
Lastly we have earlier measurements made across 
almost the entire CXB spectrum by a single telescope, namely the
{\it High Energy Astrophysics Observatory\/} ({\it HEAO\/}),
where \citet{gruber1999} (and references therein) determine the 
CXB between 3 and\,275 keV using a combination of {\it HEAO}'s Low Energy 
Detector (LED), Medium Energy Detector (MED) and High Energy Detector (HED).

Similar to the COB and CIB, there has been much interest in resolving the
sources responsible for producing the CXB, using the high angular resolution
offered by current-generation X-ray telescope. For instance, observations using
{\it XMM-Newton\/} have been able to resolve into discrete sources
nearly 100$\%$ of the low-energy end of the CXB, but the fraction drops to
about 50$\%$ at the mid range \citep{worsley2005,xue2012}, and at high
energies, observations with {\it NuSTAR\/} have only managed to resolve about
30$\%$ of the CXB \citep{harrison2016}. While detecting the X-ray emitters that
contribute to the CXB is astrophysically interesting, we have not included
these measurements in our compilation, since the absolute value of the CXB is
much more attainable in this frequency range than in the infrared to
ultraviolet range (due to much less foreground contamination).

In Fig.~\ref{cxbfig} we have plotted all of the above data. A fourth maximum 
in the CB is seen to peak at about $5 \times 10^{18}$\,Hz (20\,keV), with an
energy density that is more than two orders of magnitude below that of the
CIB and COB.  There is 
broad consensus between each CXB evaluation; however, some of the measurements 
with small uncertainties do not appear to be in full agreement with the others, particularly at lower energies where foreground subtraction is more difficult.
Presumably systematic errors (from calibration and foreground removal,
for example), would reconcile these measurements; however, we have not made
any attempt here to assess these effects, or to bring the measurements into
better agreement.  Nevertheless, it is clear that a simple thermal
Bremsstrahlung spectrum broadly fits the data -- this can be modelled as a
double power law of the form
\begin{equation}
{A\over (E/E_{\rm c})^\alpha+(E/E_{\rm c})^\beta},
\end{equation}
for specific parameters $A,E_{\rm c}, \alpha$ and $\beta$. We have taken
best-fit values for this model from \citet{ajello2008} and plotted the
resulting curve along with the data in Fig.~\ref{cxbfig}.

\subsection[Cosmic $\gamma$-ray background]{The cosmic $\gamma$-ray 
Background (CGB)\\ 
{\large [$\mathbf{\nu>10^{19}}\,$Hz, $\mathbf{\lambda<0.03}\,$nm}]}
\label{cgb}

The final portion of the CB is the CGB, which encompasses all frequencies
greater than $10^{19}$\,Hz. The primary contributors to the CGB are 
quasars/blazars \citep{inoue2014} and supernova explosions \citep{riuz2001}.
Quasars and blazars emit jets of ultra-relativistic charged particles along the 
rotational axis of their accretion disks; photons can reach enormous energies
by Compton scattering off these electrons, which we see if the jet is pointing 
close to our line of sight.  Supernovae explosions, on the other hand, 
occur when a massive star exhausts its fuel at the end of its lifetime, 
resulting in a cataclysmic core collapse and the release of a huge amount of 
energy, including $\gamma$-rays. With about 100 supernovae occurring per year
per redshift per square degree \citep{riuz2001}, their contribution to the CGB
is believed to be small compared to 
that of quasars and blazars, but not negligible.

\begin{figure}[ht!]
\begin{center}
\includegraphics[scale=0.8]{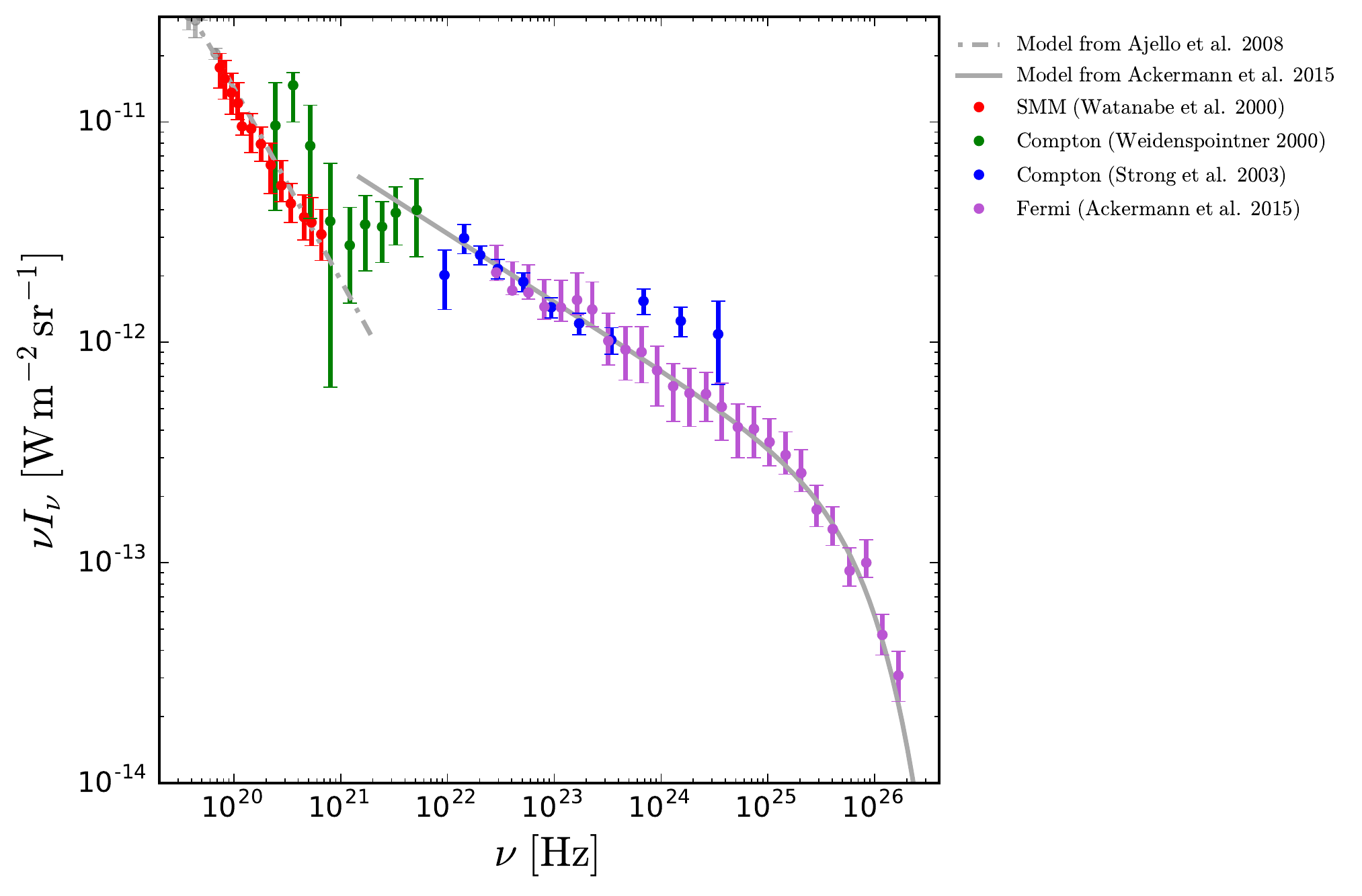} \caption[Cosmic $\gamma$-ray background]
{\label{cgbfig}
Cosmic $\gamma$-ray background radiation. There is a change in slope past 
about $10^{25}$\,Hz (40\,MeV) due to particle production from interactions 
between the high energy $\gamma$-rays and the low energy infrared photons,
which is well described by a power-law spectrum with an exponential cut-off
and plotted as a solid line \citep{ackermann2015}. The dot-dashed line shows
the model of the CXB from \citet{ajello2008}.}
\end{center}
\end{figure}

A high-energy cutoff occurs due to pair-production interactions between CGB 
photons and CIB photons \citep[this is the same effect 
used to derive CIB upper limits, see][]{dejager1992}. This cutoff is
generally thought to be around 300\,GeV, or $5 \times 10^{25}$\,Hz 
\citep{ajello2015b}, and gives a limit to the CGB -- most
higher energy photons are expected to have been converted into 
particles during 
their long intergalactic journey to the Earth. 

For this highest energy part of the CB we incorporate data from the Gamma-Ray 
Spectrometer (GRS) aboard the {\it Solar Maximum Mission\/} ({\it SMM\/})
for energies between 0.3 and 2.7\,MeV \citep{watanabe2000}, and from the
COMPton TELescope (COMPTEL) on the {\it Compton Gamma-Ray Observatory\/}
between 1 MeV and 20\,MeV \citep{weidenspointner2000}. There have also been
important measurements of the CGB from {\it Compton}'s Energetic Gamma-Ray
Experiment Telescope (EGRET), which had an effective waveband of
30\,MeV to 20\,GeV \citep{strong2004}.  {\it Compton}'s 
successor was the {\it Fermi Gamma-Ray Telescope}, which was able to make 
major improvements in the high-energy spectrum between 0.1 and 500\,GeV using 
its Large Array Telescope \citep[LAT,][]{ackermann2015}. Despite these
improvements, we still use the COMPTEL and EGRET data, since 
they bridge the gap nicely between the 2.7\,MeV {\it SMM\/} measurement and the 
0.1\,GeV {\it Fermi\/} measurement.

Figure~\ref{cgbfig} shows the CGB estimates obtained using the experiments
mentioned above.  The high-energy cutoff is clearly observed in the
{\it Fermi\/} data, since the slope 
is seen to become steeper at around $10^{25}$\,Hz (40\,GeV).  Although there
is broad agreement on the shape, not all estimates are consistent within the
uncertainties, again pointing to systematic effects between experiments.
Given {\it Fermi}'s greater resolution and ability to resolve $\gamma$-ray
point sources compared with {\it Compton}, it is 
expected to be able to isolate extragalactic sources more easily and so 
separate these out from the Galactic background signal.  In Fig.~\ref{cgbfig}
we indicate with a solid line a simple power-law model with an exponential
cutoff \citep{ackermann2015}, which approximately describes the data.

\subsection{The complete cosmic background}
\label{cb}

\begin{figure}[ht!]
\begin{center}
\includegraphics[scale=0.8]{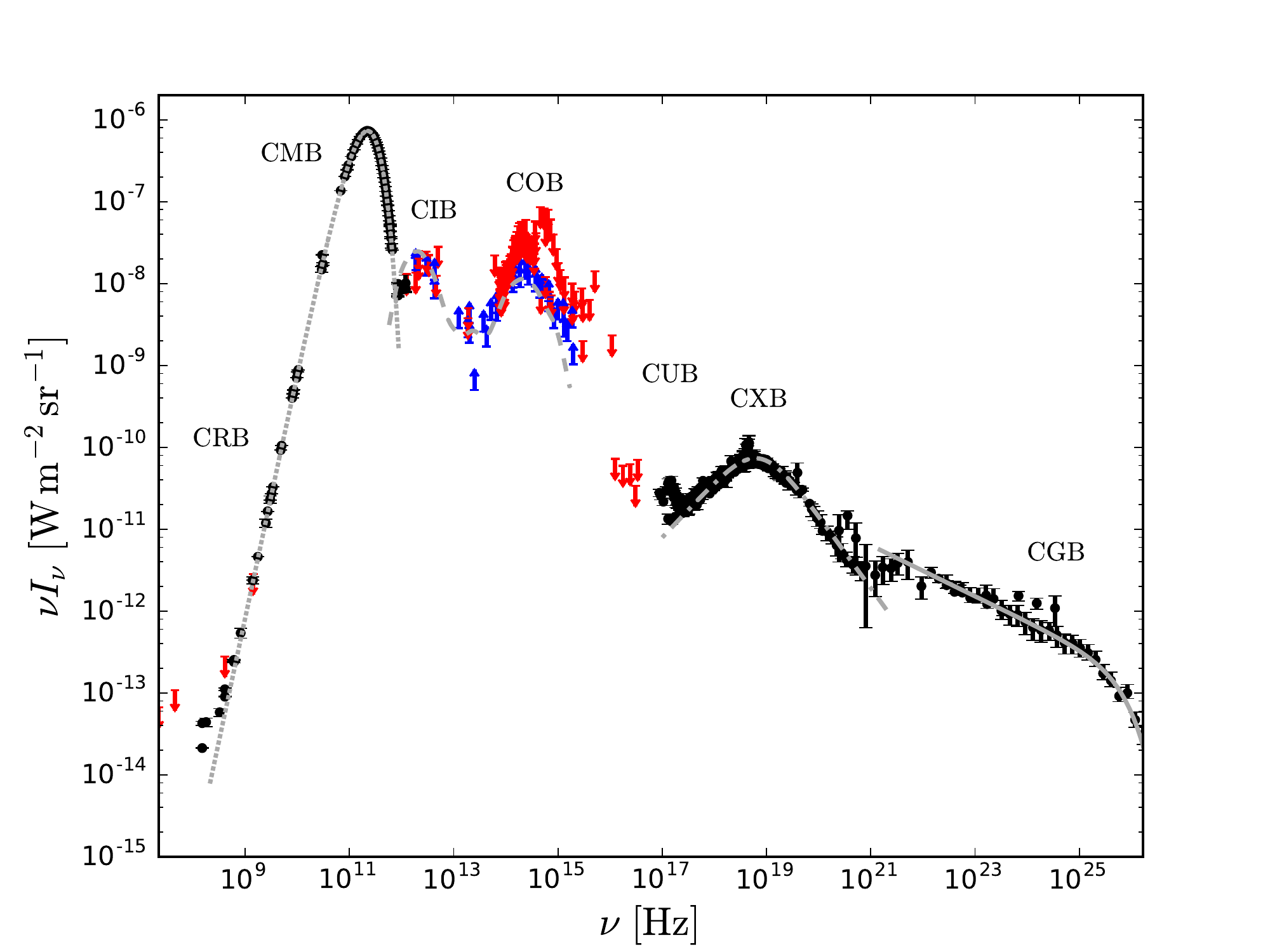}
\caption[Complete cosmic background]
{\label{cbfig}
Complete cosmic background radiation. Black points with error bars 
indicate detections, while blue and red arrows indicate lower and upper limits, 
respectively. The grey lines are models of various sections of the the cosmic
background, as outlined in the text.}
\end{center}
\end{figure}

We can now piece together the cosmic background radiation over 
the entire electromagnetic spectrum. Combining Figs.~\ref{crbfig} 
to~\ref{cgbfig} gives the overall result, shown in Fig.~\ref{cbfig}.  This 
represents the energy distribution of photons permeating the Universe, and 
includes radiation emitted from sources over the whole of cosmic history.

\begin{figure}[ht!]
\begin{center}
\includegraphics[scale=0.8]{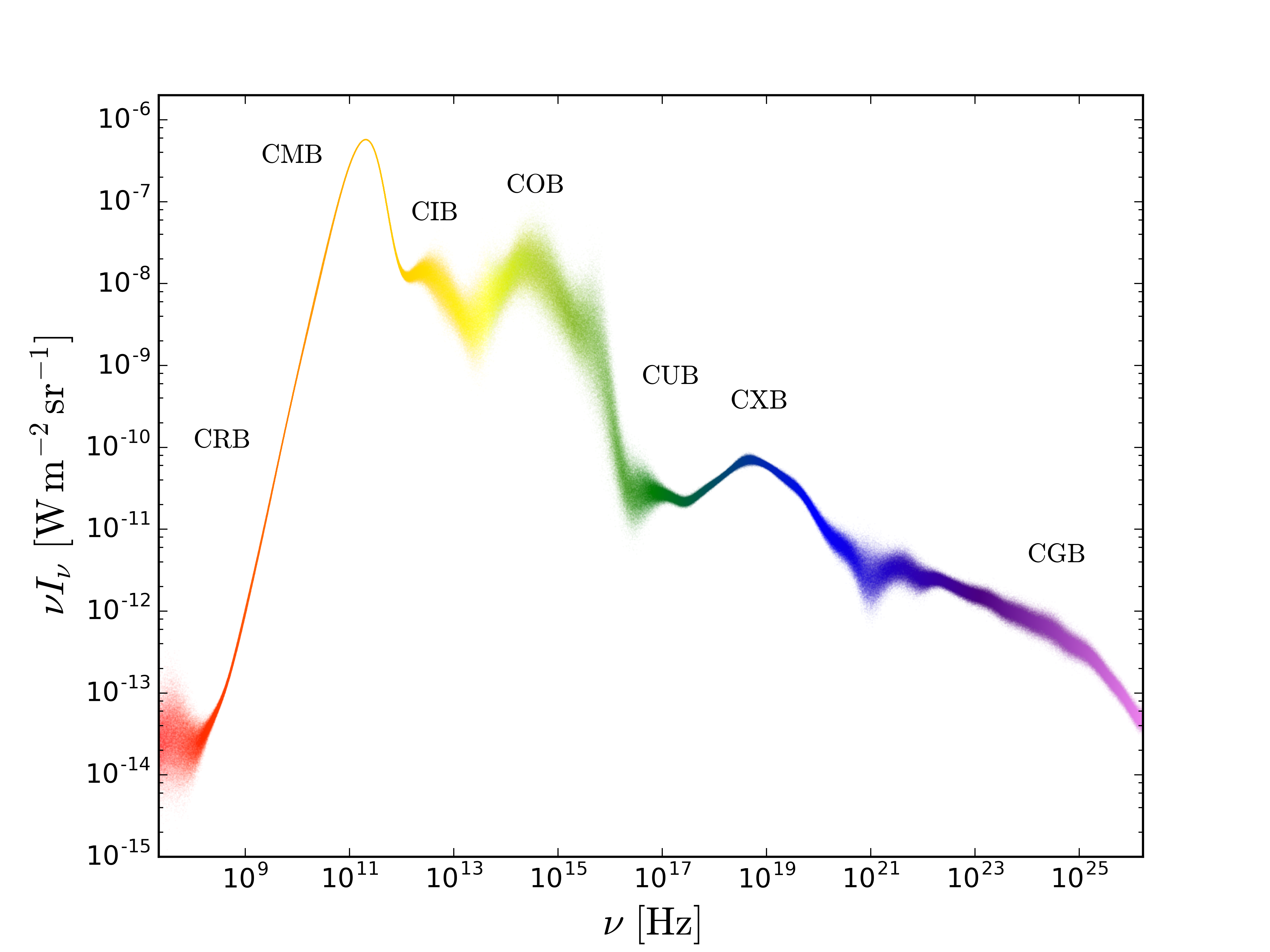}
\caption[Complete cosmic background]
{\label{splatter}
Overall estimate for the continuous cosmic background radiation. The data in 
Fig.~\ref{cbfig} were used to derive upper and lower limits at all 
frequencies, which is proportional to the spread in this ``splatter plot''.
Where the background is well measured the line is narrow, and it becomes
thicker in regions where the uncertainties are larger.}
\end{center}
\end{figure}

As an alternative way of presenting the data,
we have produced a best-guess continuous curve from 
the discrete measurements of Fig.~\ref{cbfig}. The thickness of the line here 
is proportional to the uncertainties. These were estimated by 
extrapolating and smoothing a set of continuous curves along the upper and 
lower uncertainty values of each data point in Fig.~\ref{cbfig} (if only an 
upper/lower limit is present, this is taken to be the upper/lower uncertainty 
value), and letting the width be $\approx 2 \sigma$, corresponding to upper and 
lower two-sided 95$\%$ confidence interval limits. This procedure was used to 
determine the spread between the two curves, which is shown as a ``splatter 
plot'' in Fig.~\ref{splatter}.  Where the curve is narrow, the spectrum is
well constrained, but where the spread is wide there is still lots of room
for improving the measurements.

\section{Line emission perturbations}
\label{lineemissionperturbations}

So far every emission process we have discussed has been spectrally continuous
-- synchrotron radiation, blackbody radiation, thermal bremsstrahlung 
radiation and Compton processes each depend smoothly on the energies of 
the particles involved. However, line emission (essentially Dirac delta 
functions in the cosmological rest frame) are also present in the CB\@. 
If we could extract these discrete features, this would provide information 
not only about physical conditions projected along the line of sight, but also 
about the distribution of redshifts (and therefore distances) of the sources
of emission. The technique of detecting line 
emission at different frequencies over large solid angles is known as 
``intensity mapping''
\citep[e.g.,][]{hogan1979,scott1990,madau1997,suginohara1999,chang2008,mao2008,kovetz2017}, 
and can be viewed as a way to 3-dimensionally map {\it volumes\/} of space. The 
sensitivity required to successfully glean statistical information on 
cosmological scales from intensity mapping is quite modest, but the most challenging aspect of these experiments is the removal of foreground emission and systematic effects. Experimental results are 
only recently beginning to appear, as experiments are being developed to
overcome these challenges.
In this section we discuss the effects on the CB of 
a few of the most cosmologically important lines by estimating their 
monopole terms as functions of redshift, and hence their contribution to the
spectrum of the monopole.  We will also present some recent 
experimental attempts to constrain these line contributions to the 
background sky.

\subsection{The H\textsc{i} line}
\label{h1}

The spin-flip transition in neutral hydrogen, H\textsc{i}, also known 
as the 21-cm line, has a rest-frame frequency of 1420\,MHz, corresponding to
0.00587\,eV\@. It comes from the transition between the aligned 
(triplet) and anti-aligned (singlet) electron-proton spin states, the latter
of which has slightly lower energy.  Although intrinsically weak (this is a
forbidden transition, with a lifetime of around 10 million years), the 
fact that hydrogen is the most abundant element in the Universe makes 
this line an important tracer of neutral gas everywhere, including for
cosmological large-scale structure.
There is currently much promise for the 21-cm 
line to probe the epoch of reionization (EoR) at $z\sim10$, which marks the end of 
the cosmic ``dark ages'', where we currently only have indirect information 
\citep[see][for a review of the prospects for 21-cm intensity 
mapping]{furlanetto2006b}.

Prior to the EoR, the intensity of the 21-cm mean background
depends largely on the hydrogen spin temperature, 
$T_{\mathrm{s}}$, a measure of the ratio of atoms in the singlet versus 
triplet configurations:
\begin{equation}
\label{densityratio}
\frac{n_{1}}{n_{0}} = 3 e^{- \Delta 
T_{\mathrm{H\textsc{i}}}/T_{\mathrm{s}}}.
\end{equation} 
Here $\Delta T_{\mathrm{H\textsc{i}}} = 
\Delta E_{\mathrm{H\textsc{i}}}/k = 0.068$\,K is the transition 
energy in temperature units. Also necessary for calculating the line 
intensity is the mean fraction of neutral hydrogen, 
$\bar{x}_{\mathrm{H\textsc{i}}}$, since only bound hydrogen atoms emit 
21-cm radiation. The evolution of the neutral fraction is particularly
important following the EoR, when hydrogen transitioned from being
mostly neutral to its present state of being mostly ionized.
Expressing the intensity as a CMB-subtracted brightness 
temperature (which can be converted to SI units through
Eq.~\ref{temperaturetoInu}), we have  \citep{furlanetto2006a}
\begin{equation}
\label{H1temp}
T_{\mathrm{H\textsc{i}}} \approx 27 \bar{x}_{\mathrm{H\textsc{i}}} 
\left( \frac{1+z}{10} \right) ^{1/2} \left( 1 - 
\frac{T_{\gamma}}{T_{\mathrm{s}}} \right) \mathrm{mK},
\end{equation}
where $T_{\gamma} = (1+z) T_{\rm CMB}$, and the factor of 27 arises from 
cosmological parameters within the best-fit $\Lambda$CDM model, combined
with the properties of the $\mathrm{H\textsc{i}}$ line.

\begin{figure}[ht!]
\begin{center}
\includegraphics[scale=0.7]{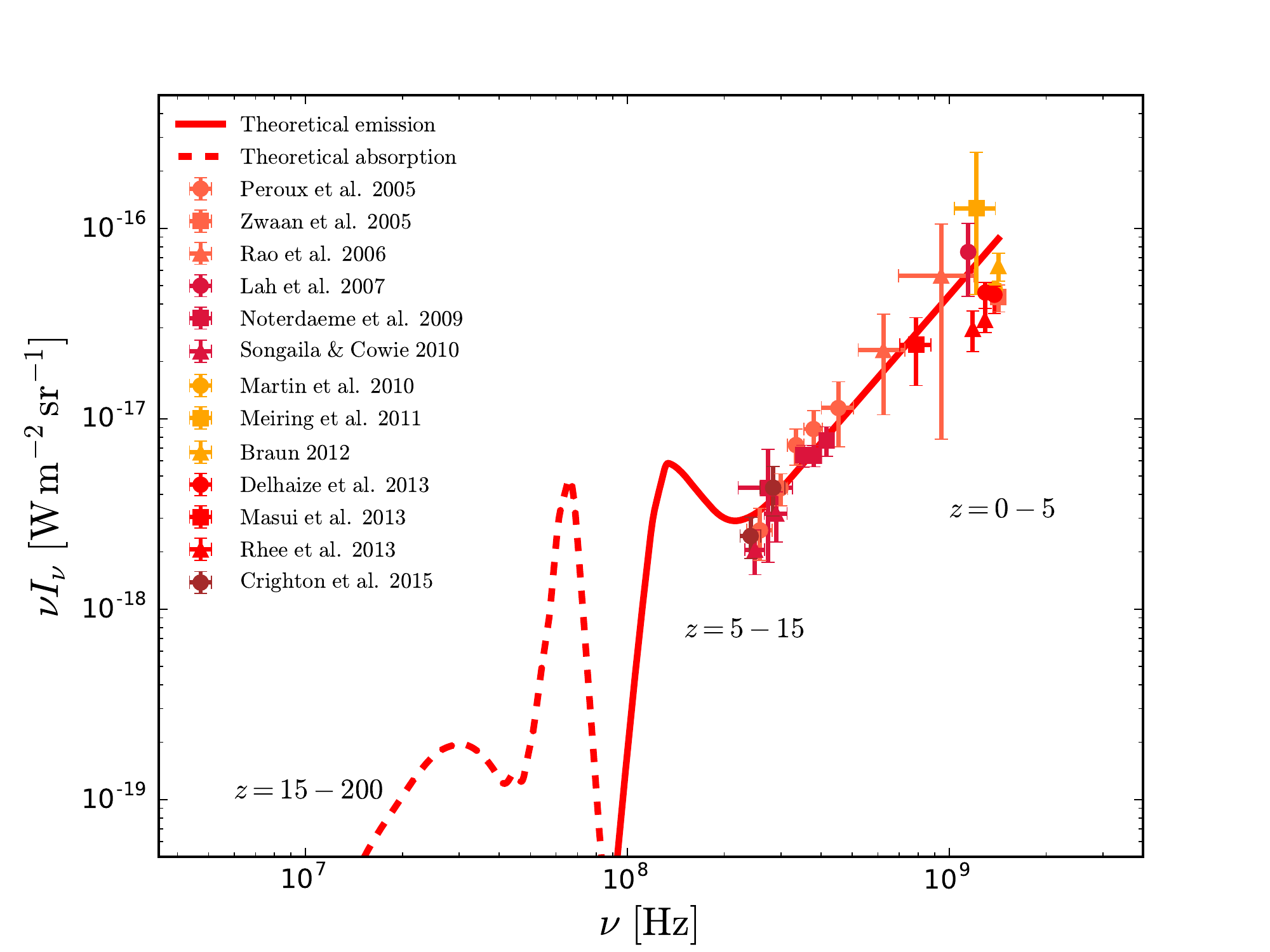}
\caption{\label{H1fig}
H\textsc{i} 21-cm line contribution to the CB monopole.  The red solid curve is
the predicted intensity of line emission at redshifts up to redshifts of about
15 and the epoch of reionization, after which the red dashed curve represents
absorption of the background CMB spectrum during the dark ages when the
Universe was neutral. The points with error bars are various measurements of
the H\textsc{i} line strength, which go out to a redshift of about 5.}
\end{center}
\end{figure}

Models of
the evolution of the gas density and temperature, including ionization sources
in the early Universe, can be used to calculate the spin 
temperature and ionization fraction as a function of redshift
\citep[e.g.,][]{pritchard2008,mirocha2013,dave2013,rahmati2013,barnes2014,bird2014,yajima2015,ghara2015,padmanabhan2017}.
Here we have taken the results of several representative studies
\citep{furlanetto2006a,pritchard2008,mirocha2013,yajima2015,ghara2015} 
between $z = 0$ and $z = 200$ and produced an average spectrum, shown 
in Fig.~\ref{H1fig}; this gives merely a rough estimate
of the integrated 21-cm line intensity, and hence we have not displayed 
any uncertainty. At low frequencies (coming from epochs above $z = 15$),
the 21-cm line is seen in 
absorption against the CMB, instead of emission, because the hydrogen 
gas adiabatically cools below the CMB temperature. We have plotted the 
absolute value of this negative temperature difference as the dashed line in 
Fig.~\ref{H1fig}.

We now turn our attention to observations. The background 21-cm 
brightness temperature is difficult to measure directly, since it is hard
to absolutely calibrate most radio detectors, and there are also
continuum foregrounds that will contaminate the measurements.  However, the 
21-cm monopole can be determined indirectly by estimating 
$\Omega_{\mathrm{H\textsc{i}}}$, the neutral hydrogen density in units of
the cosmological critical density, performed via a census of
hydrogen absorption (or emission)
lines. In the limit where $T_{\gamma} \ll T_{\mathrm{s}}$ (applicable 
when $z \lesssim 5$) Eq.~(\ref{H1temp}) can be simplified and 
$T_{\mathrm{H\textsc{i}}}$ estimated in terms of 
$\Omega_{\mathrm{H\textsc{i}}}$.  Another approach is to cross-correlate noisy
radio maps of 21-cm emission with 3D maps of galaxies. This measures the extent
to which ${\mathrm{H\textsc{i}}}$ clusters with galaxies, rather than measuring
the $\Omega_{\mathrm{H\textsc{i}}}$ or the 21-cm monopole directly. When can
then use our understanding of structure formation to infer the hydrogen
abundance.

Such measurements of $\Omega_{\mathrm{H\textsc{i}}}$ have been carried out
using radio surveys \citep{zwaan2005,lah2007,martin2010,braun2012,rhee2013,delhaize2013,masui2013} and optical surveys \citep{peroux2005,rao2006,noterdaeme2009,songalia2010,meiring2011,crighton2015}. Over the redshift range of $z=0$--5, 
$T_{\mathrm{H\textsc{i}}}$ has been found to be of the order
100\,$\mu$K (roughly $10^{-17}$--$10^{-16} \, \mathrm{W \, m^{-2} \, 
sr^{-1}}$ at these frequencies). We show these estimates in 
Fig.~\ref{H1fig}, along with the theoretical calculations discussed 
above. Future experiments promise to probe the ``step'' expected at around 
100\,MHz, whose detailed profile will tell us about the processes that
reionized the Universe.

\subsection{Lines from early Universe chemistry}
\label{EUlines}

In the standard picture of structure formation, there are no important
atomic processes happening between the time of cosmological recombination
at a redshift of about 1000 (when the Universe was a few hundred thousand
years old) and when the first objects collapsed and turned into stars
at redshifts perhaps around 15 (when the Universe was a few hundred million
years old).  The formation of the first stars began the process of creation
of the heavy elements that make the
world so interesting today.  However, chemical processes started before that,
operating on the hydrogen, helium and trace amounts of lithium and other
light elements that were produced primordially \citep[see e.g.,][]{GalliPalla}.
Molecular hydrogen was able to form via intermediary species such as
H$_2^+$, H$^-$, and HeH$^+$, and was a major coolant for the formation of
the first structures.  Detection of lines from this early H$_2$-cooling phase
would be an important probe of the early Universe, but will be extremely
challenging \citep[although there is hope through future space missions,
such as the {\it Origins Space Telescope},][]{OST2016}.

Additionally there will be trace
amounts of other molecules, such as HD and LiH, and there have been suggestions
that these may be important at certain epochs.  Despite lithium having very
low abundance in the early Universe, at one point it was proposed
\citep{Loeb2001,Stancil2002} that absorption of the CB by the (redshifted)
$6708\,\AA$ line of neutral Li might be important.  However, it was shown
\citep{Switzer2005}
that the small excess of UV photons created during hydrogen recombination
is enough to keep lithium ionized during these dark ages, and hence the
line is cosmologically unimportant.  In a similar way, the possibility of
free-free absorption by H$^-$ in the early Universe \citep{Black2006} was
substantially revised down to negligible levels \citep{Schleicher2008}.
There may still be scope for early Universe chemistry to have observational
consequences in the distant future, but in the meantime attention on line
processes is focused on the more accessible lower redshift regime.

\subsection{The C\textsc{ii} line}
\label{C2}

There are observable IR fine-structure lines emitted from 
elements such as carbon, nitrogen and oxygen that are important for 
cooling in star-forming regions of galaxies. Modelling these line 
emission processes is more complicated than for H\textsc{i},
with many possible spin 
transitions from not only neutral atoms but ionized atoms as well, and 
with radiative transfer effects 
needing to be included to account for the diverse physical conditions.
Nevertheless the general approach to 
estimating the background contribution remains more or less the same. 
As a specific example, we will first focus on emission from 
C\textsc{ii}, singly ionized carbon.

Since carbon is produced in stars and the $^{2}P_{3/2}$--$^{2}P_{1/2}$
transition temperature (91\,K) matches that of photo-dissociation
regions near to bright stars, C\textsc{ii} plays an 
important role in cooling \citep{dalgarno1972}.  In fact it can be the
brightest line in galaxy spectra, containing up to 0.1$\%$ of the bolometric 
luminosity \citep{crawford1985}. This fine-structure line is
at 157.7\,$\mu$m (or 1900\,MHz), placing it in the far-IR\@.

Two approaches have been developed for modelling this transition in the
cosmological context. The first 
follows directly from the discussion of H\textsc{i}, calculating the 
spin temperature as a function of redshift through 
Eq.~(\ref{H1temp}) \citep{gong2012,kusakabe2012}. The second method, outlined 
in \citet{visbal2010}, \citet{uzgil2014} and \citet{silva2015}, is simpler,
and uses an integral over emissive sources assuming no absorption along
the line of sight, using the following equation:
\begin{equation}
\label{C2intensity}
I_{\nu, \mathrm{C\textsc{ii}}} = \frac{1}{4 \pi} \int_{0}^{\infty} 
\frac{\epsilon_{\mathrm{C\textsc{ii}}}(\nu_{\mathrm{C\textsc{ii}}})}{(1+
z)^{3}} \frac{dl}{dz} dz.
\end{equation}
Here the line element $dl/dz = c/(1+z)H(z)$ contains the cosmology,
and $\epsilon_{\mathrm{C\textsc{ii}}}$ is the proper volume emissivity of 
C\textsc{ii} radiation at rest frequency $\nu_{\mathrm{C\textsc{ii}}}$.
If all C\textsc{ii} emission comes from gas in
galaxies located in dark matter halos, we can write the proper volume 
emissivity as
\begin{equation}
\label{C2density}
\epsilon_{\mathrm{C\textsc{ii}}}(\nu) = 
\int_{M_{\mathrm{min}}}^{\infty} L_{\mathrm{C\textsc{ii}}}(\nu,M) 
\frac{dn}{dM} dM,
\end{equation}
where $dn/dM$ is the dark matter halo mass function
\citep[e.g.,][]{tinker2008}, 
$M_{\mathrm{min}}$ is the minimum halo mass required to form a galaxy 
and $L_{\mathrm{C\textsc{ii}}}(\nu,M)$ is the C\textsc{ii} luminosity 
at frequency $\nu$ for a dark matter halo of mass $M$. We then make the 
further simplification that the C\textsc{ii} luminosity of a galaxy is 
proportional to the star-formation rate,
\begin{equation}
\label{C2luminosity}
L(\nu,M) = R_{\mathrm{C\textsc{ii}}} \dot{M} \delta(\nu - 
\nu_{\mathrm{C\textsc{ii}}}),
\end{equation}
where the constant $R_{\mathrm{C\textsc{ii}}}$ can be calibrated through
observations of nearby galaxies and has the typical value (in solar units) of
$R_{\mathrm{C\textsc{ii}}}\approx6\times10^{6}\,\mathrm{L_{\odot} 
\, M_{\odot}^{-1} \, yr}$ \citep{righi2008}, and $\dot{M}$ is the
star-formation rate. The delta function
$\delta(\nu - \nu_{\mathrm{C\textsc{ii}}})$ here ensures that the line
emission occurs at the proper frequency.

For $\dot{M}$, we can write
\begin{equation}
\label{mdot}
\dot{M} = \frac{f_{\ast}}{t_{\rm s}} \frac{\Omega_{\rm b}}{\Omega_{\rm m}} M,
\end{equation}
where $\Omega_{\rm b}$ and $\Omega_{\rm m}$ are the baryonic and matter density 
fractions of the Universe and $f_{\ast}$ is the fraction of baryons contained 
in stars, which form on a typical timescale $t_{\rm s}$. 
Integrating the volume emissivity over redshift then yields
\begin{equation}
\label{C2intensity2}
I_{\nu,\mathrm{C\textsc{ii}}}(\nu) = \frac{c}{4 \pi} 
\frac{1}{\nu_{\mathrm{C\textsc{ii}}} H(z)} R_{\mathrm{C\textsc{ii}}} 
\frac{f_{\ast}}{t_{\rm s}} 
\frac{\Omega_{\rm b}}{\Omega_{\rm m}}\int_{M_{\rm min}}^{\infty} M \frac{dn}{dM} dM.
\end{equation}
From this equation it can be seen that one generally has to estimate $f_{\ast}$
and $t_{\rm s}$, which could in principle depend on the mass and lead to more
complicated models. Improvements can also be considered, including
treating heavy element 
abundance \citep{yue2015}, and having $R_{\mathrm{C\textsc{ii}}}$ depend on
galaxy type, but these do not change the order of magnitude of the result.

In Fig.~\ref{fslinesfig}, we show an average of the resulting models from
\citet{visbal2010}, \citet{gong2012}, \citet{kusakabe2012}, \citet{uzgil2014}
and \citet{silva2015}, giving approximately the expected magnitude of the
C\textsc{ii} signal as it varies with redshift.

\subsection{Other fine-structure lines}
\label{otherlines}

Other far-IR lines from low-ionization states of common elements are also
prominent in galaxy spectra.
To gauge the contribution to the CB from other major cooling 
lines, we can use Eq.~(\ref{C2intensity2}) to scale the cosmic 
C\textsc{ii} monopole spectrum, since it just depends on $R$ and $\nu$
for each line. 
\citet{righi2008} provides $R$ ratios for two C\textsc{i} transitions 
(610 and 371\,$\mu$m), two O\textsc{i} transitions (145 and 
63\,$\mu$m), two O\textsc{iii} transitions (88 and 52\,$\mu$m), two 
N\textsc{ii} transitions (122 and 205\,$\mu$m) and one N\textsc{iii} 
transition (57\,$\mu$m). We plot these line intensities alongside that 
of C\textsc{ii} in Fig.~\ref{fslinesfig}. Other studies have 
approached the calculation in different ways 
\citep[e.g.,][]{kusakabe2012}, but give generally similar results.

\begin{figure}[ht!]
\begin{center}

\includegraphics[scale=0.7]{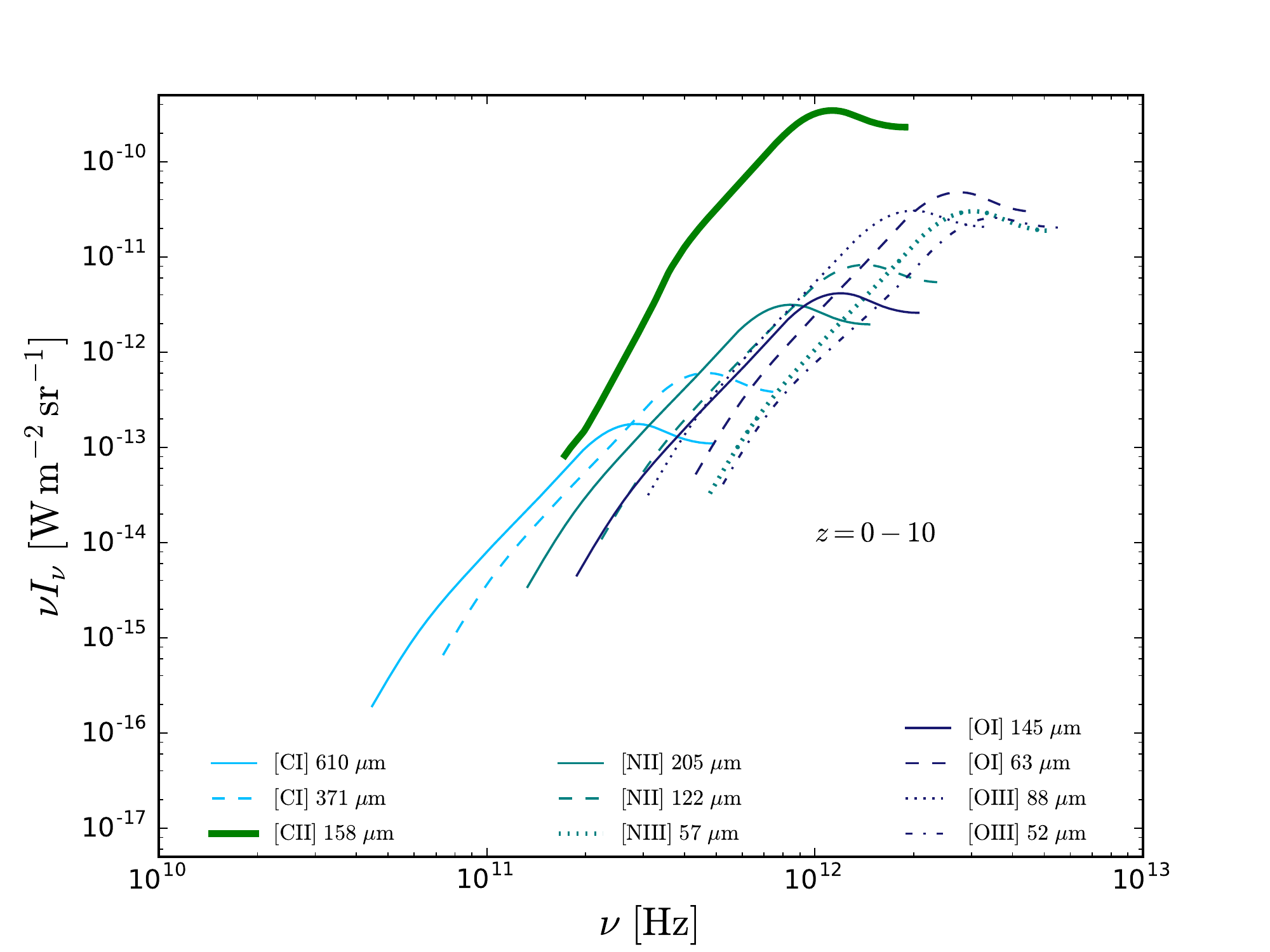}
\caption{\label{fslinesfig}
Expected contribution to the CB from fine-structure line emission, showing the
redshift range $z=0$ to $z=10$. The C\textsc{ii} line intensity, shown as the
thick green line, has been estimated by averaging calculations from 
\citet{visbal2010}, \citet{gong2012}, \citet{kusakabe2012}, \citet{uzgil2014}
and \citet{silva2015}. The other lines have been scaled from each line
frequency, along with ratios of the line luminosity to 
star-formation rate, according to Eq.~(\ref{C2intensity2}).}
\label{fslinesfig}
\end{center}
\end{figure}

\subsection{CO lines}
\label{CO}

The most abundant molecule in the Universe is $\mathrm{H_{2}}$, but due 
to the lack of dipole transitions it is hard to detect. The predominant 
tracer of molecular gas in galaxies is then carbon monoxide 
\citep[see][for a review of carbon monoxide and $\mathrm{H_{2}}$ in 
star-forming galaxies]{bolatto2013}. The main lines are
rotational-vibrational
transitions from the total angular momentum state $J$ to $J-1$, and the photons emitted occur
in the millimetre and
submillimetre bands at multiples of 115\,GHz.

Several studies have estimated the global CO emission using a similar 
approach to Section~\ref{C2}, by scaling the IR galaxy luminosity 
\citep{lidz2011,carilli2011,pullen2013,li2015} or the star-formation 
rate \citep{breysse2014}. Alternatively, one can use a line 
luminosity to star-formation rate ratio, but model the star-formation 
rate through the dark matter halo abundance \citep{righi2008,padmanabhan2018}, or use the 
output of large-scale galaxy formation simulations \citep{gong2011}. 
There is general agreement that the expected signal is in the $\mu$K regime,
although predictions vary in detail. We 
have taken an average of the above models and calculated the intensity 
out to $z \approx 10$, which we show in Fig.~\ref{COfig}. 

This estimate can be extended out to the higher CO transitions by
scaling $R_{J}$ to different values of $J$
\citep[from e.g.,][]{righi2008,visbal2010} and scaling the rest frequency of
each line. We include these curves in Fig.~\ref{COfig} up to $J=13$,
along with the sum of all CO transition intensities.

\begin{figure}[ht!]
\begin{center}
\includegraphics[scale=0.7]{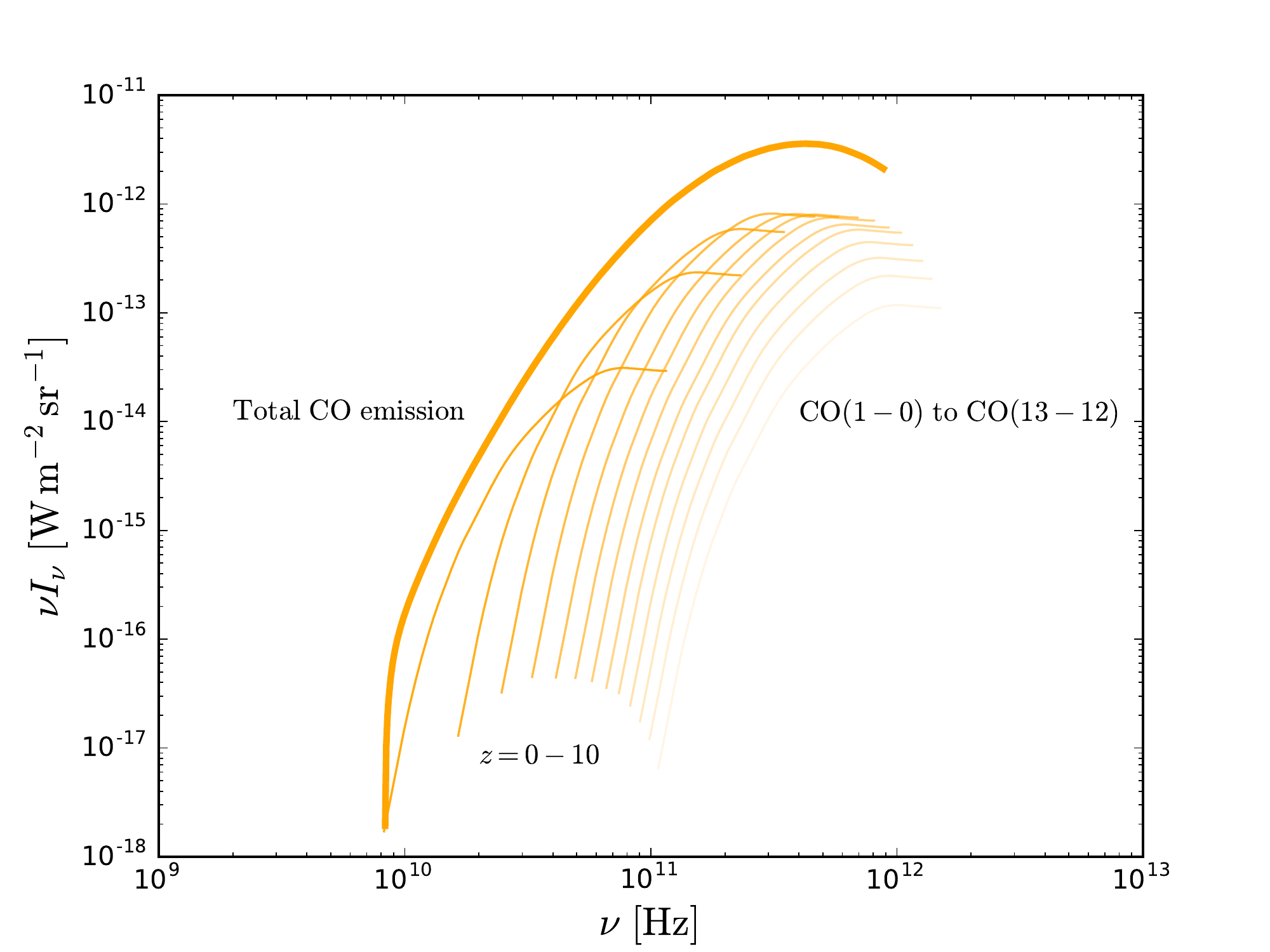}
\caption{\label{COfig}
CO monopole intensity model 
\citep[based on calculations from][]{righi2008,gong2011,lidz2011,carilli2011,pullen2013,breysse2014,li2015,padmanabhan2018},
shown as the thick light orange line. Contributions from individual CO
rotational-vibrational transitions are shown as thinning orange lines from
$J=1$ to 0, up to a maximum transition between $J=13$ and 12.}
\label{COfig}
\end{center}
\end{figure}

\subsection{The Ly$\alpha$ line}
\label{lyalpha}

As well as radio and far-IR lines, the spectra of galaxies also have
prominent emission lines at visible wavelengths.  Of particular interest
is the Lyman\,$\alpha$ line at $122\,$nm (or $2.47\times10^{15}$Hz),
which results from a transition from the $n=2$ to $n=1$ state in hydrogen.
This line plays an important role in the opacity of the ISM because of the
abundance of hydrogen and the supply of ultraviolet continuum photons from hot
young stars.

Theoretical intensity models have been constructed 
\citep[e.g.,][]{pullen2013,silva2013,gong2014}, taking into account 
radiative and collisional recombinations, with stars being the sources of
radiation. In contrast to CO and C\textsc{ii}, which are generally
confined to the inner parts of galaxies, hydrogen 
emission also comes from diffuse gas in galaxy halos and to some extent
from the intergalactic medium. Detailed calculations involve modelling 
radiative transfer in clumpy and dusty regions, with a great deal of
uncertainty in the specifics.
We merely display an average of representative results in 
Fig.~\ref{lyalphafig} for redshifts up to $z \approx 10$.

Ly\,$\alpha$ emission can also be estimated empirically.  For example
\citet{croft2016} recently measured the brightness \emph{fluctuations} 
of the Ly$\alpha$ emission by cross-correlating optical spectra with 
galaxies tracing the large-scale structure of the Universe. This 
measures the product of the mean Ly$\alpha$ intensity and 
linear bias factor $b_{\alpha}$ (parameterizing
the cosmic large-scale clumpiness of emission), and it was found that 
$I_{\nu} = (1.1 \pm 0.3)\times10^{-24}\,{\rm W}\,{\rm m}^{-2}\,{\rm sr}^{-1}$
at $z \approx 2$, assuming $b_{\alpha}=3$. 
We have included this measurement in Fig.~\ref{lyalphafig}.
Although the overall bias is uncertain, for any reasonable value this 
result is roughly an order of magnitude larger than the theoretical 
estimates. The extra emission could come from an unresolved 
population of Ly$\alpha$ emitters, which have not been taken into account
in the models, or could be due to a number of possible systematic errors.
In summary, in this redshift range the total Ly$\alpha$ brightness is highly
uncertain, both in terms of theoretical models and empirically. Decreasing this
uncertainty is currently an area of active research in the community
studying galaxy evolution.

Another cosmologically relevant perturbation to the CB spectrum
resulting from hydrogen (and, to a lesser extent, helium) is from the 
high redshift recombination (or ``combination'', since the atoms had never
previously been neutral) process, occurring at redshift
$z \approx 1100$. Numerical calculations have shown that this
produces a distortion to the CMB of amplitude $\Delta I_{\nu} / I_{\nu} \sim 
10^{-7}$ \citep{dubrovich2004,kholupenko2005,wong2006,chluba2007}. 
Recombination contributes photons not only from the Ly$\alpha$ 
transition, but from {\it all\/} possible hydrogen and helium 
transitions, including 2-photon processes. We show in Fig.~\ref{lyalphafig}
the specific results from \citet{chluba2007}, and note that the 
Ly$\alpha$ transition discussed above corresponds to the highest peak,
while the remaining peaks come from other hydrogen transitions.
The detection of these spectral deviations will be extremely challenging,
but there are attempts being discussed, for instance by a future mission like
PIXIE \citep{kogut2011}.

\begin{figure}[ht!]
\begin{center}
\includegraphics[scale=0.7]{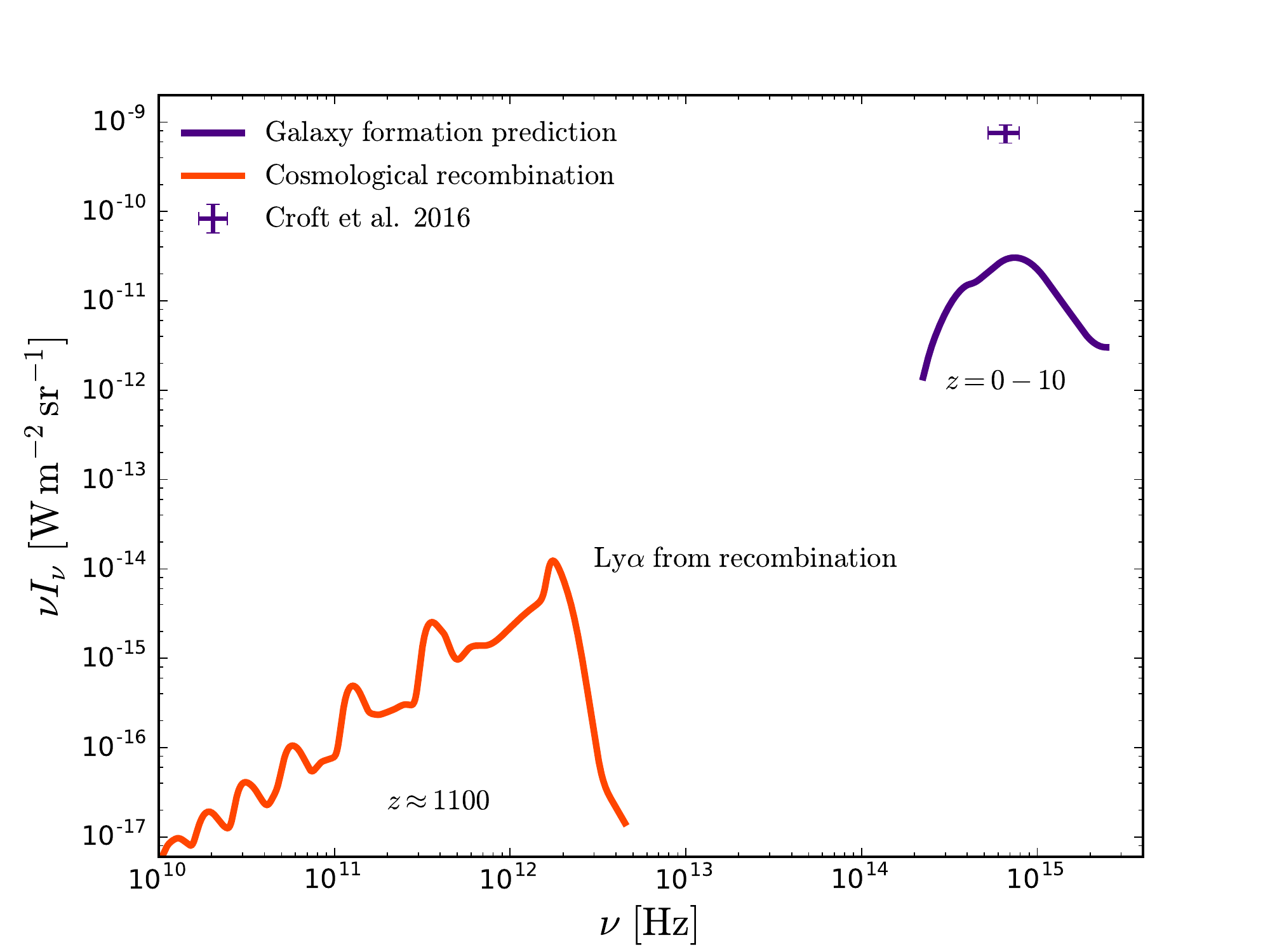}
\caption{\label{lyalphafig}
Prediction of the Ly$\alpha$ contributions to the total CB 
\citep[see][]{pullen2013,silva2013,gong2014}.  The solid purple curve 
shows a theoretical estimate from structure out
to a redshift of 10. The observationally-derived estimate of \citet{croft2016}
is shown as the point with error bars; the theoretical predictions are highly
uncertain and the measurement may be contaminated by systematic effects, either
or both of which may explain the discrepancy.  Also shown in dark orange is 
the contribution from cosmological recombination at $z \approx 1100$ 
\citep{chluba2007} from the phase transition in the early Universe, when
hydrogen first became neutral. The highest peak in this curve is from
Ly$\alpha$, with others peaks coming from different hydrogen transitions.}
\label{lyalphafig}
\end{center}
\end{figure}

\subsection{The Fe X-ray line}
\label{feline}

The last line we will consider is the 6.4-keV X-ray line from iron atoms,
sometimes called the K$\alpha$ line.  This is seen in AGN accretion disk 
spectra, being the brightest and most prominent spectral
feature \citep[e.g.,][]{nandra1989,pounds1989,fukazawa2011}. In this process,
iron atoms become ionized by losing an inner shell electron, and the gap is 
subsequently filled by an outer electron, emitting a 6.4\,keV, or $1.55 
\times 10^{18}$\,Hz, photon. Intensity mapping of this particular line 
has been suggested as a tool for investigating galaxy evolution by 
probing the power spectrum of highly redshifted AGN in a large-area survey
\citep{hutsi2012,kolodzig2013}. 

Continuing with our simple modelling approach, we can estimate the 
contribution to the sky brightness of the Fe line.  Note that this 
should be considered as an upper limit, since in practice not {\it every\/}
dark matter halo will host an AGN \citep{akylas2012,ueda2014}. To do this,
we again want to scale Eq.~(\ref{C2intensity2}). X-ray iron line luminosity 
measurements from AGN \citep{ricci2014} can be used to estimate $L_{\rm Fe}$
to be about $2\times 10^{34}$\,W, and combining this with typical AGN
star-formation rates of around $10$\,M$_{\odot}$\,yr$^{-1}$
\citep{mullaney2015} yields
$R_{\mathrm{Fe}}\approx5\times10^{6}\,{\rm L}_\odot\,{\rm M}_\odot^{-1}
\,{\rm yr}$. The resulting scaled spectrum is plotted in 
Fig.~\ref{totallinefig}, with downward arrows suggesting that it should
be considered as an upper limit only.

\subsection{Complete line background}
\label{completelinebackground}

Figure~\ref{totallinefig} combines each line spectrum presented in this
section.  These are of course not the only line emission processes 
occurring in the Universe, but can be considered to be the most luminous or the most abundant. In comparison with the complete CB, these line emission features lie far below the continuum background, and it would be very hard to detect any of these features in the CB 
monopole. Despite this, there is a real prospect of being able to see 
the spatially fluctuating part of some of these brighter lines through future 
intensity mapping experiments.

\begin{figure}[ht!]
\begin{center}
\includegraphics[scale=0.7]{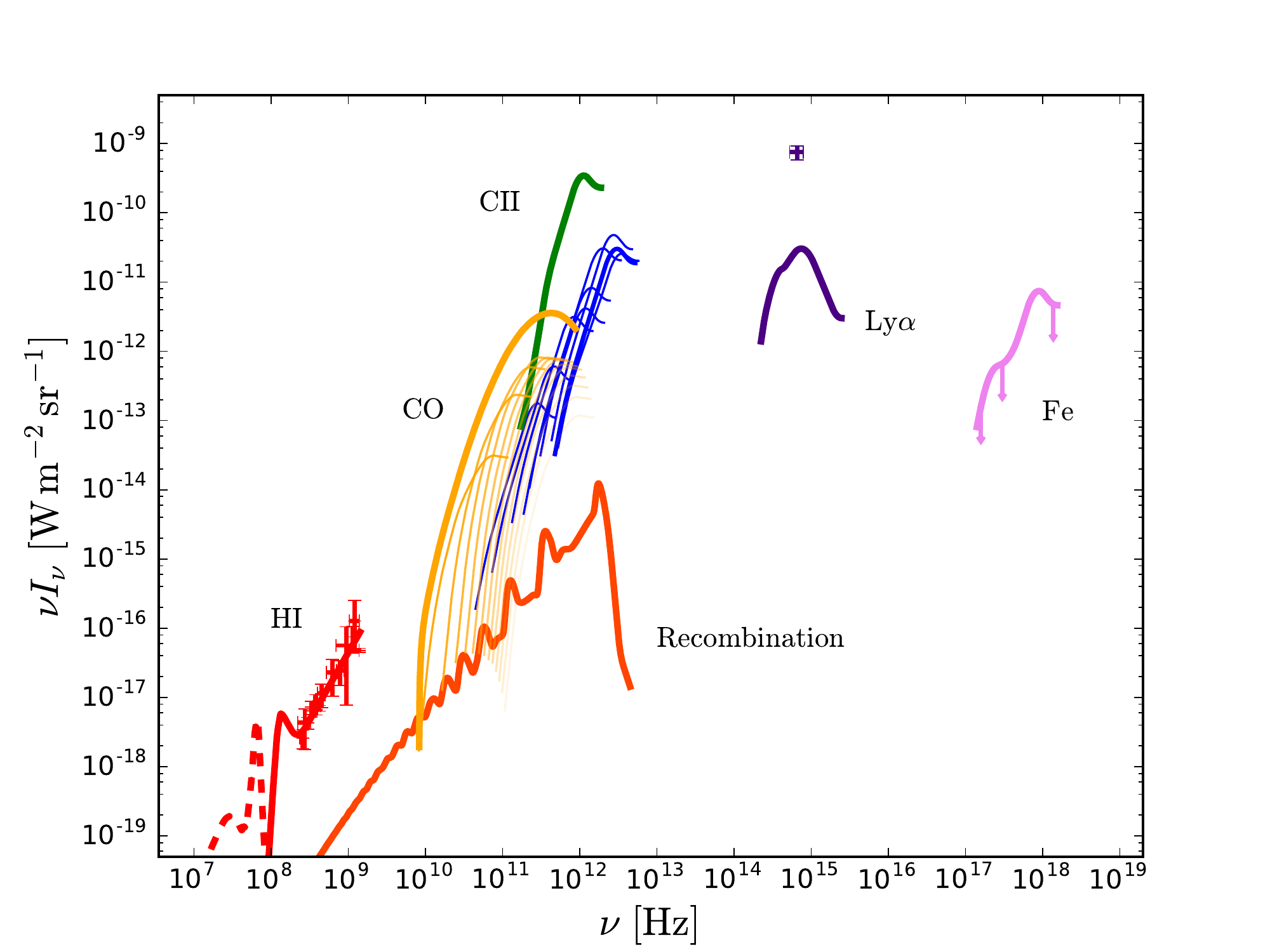}
\caption{\label{totallinefig}
Multiple line contributions to the total CB\@. Highlighted are 
H\textsc{i} (red), recombination radiation (dark orange),
CO (light orange), C\textsc{ii} (green),
other C, O and N fine-structure lines (blue), Ly$\alpha$ 
(purple), and Fe (pink). Solid lines are predictions from models, while error
bars indicate measurements. Note that all of these line emission processes are
orders of magnitude smaller than the continuum that dominates the background.}
\label{totallinefig}
\end{center}
\end{figure}

\section{Beyond the monopole}
\label{beyondthemonopole}

We have focused here on the average sky brightness at each frequency, i.e., the
spectrum of the cosmic monopole. This is proportional to the energy density as 
a function of frequency, and hence provides important information on the 
census of energy processes that occur through the history of our Universe. An 
obvious next step would be to characterize the spectrum of the cosmic dipole. 
In practice this is dominated by the fact that we are moving relative to a 
frame defined by radiation on very large scales. Variations in the dipole 
caused by there being more structure on one side of the sky than another are 
subdominant to this relative-frame ``boosting'' effect,
hence the spectrum of the cosmic dipole is essentially the same as the
monopole \citep[just lower by a factor of about $10^{-3}$ because
$v/c \approx 10^{-3}$ for our motion,][]{ScottSmoot2016,Burigana2017}.
Hence measurement of the spectrum of the dipole pattern is not likely to
provide much new information about the Universe. However, the same is not
true about sky variations on smaller angular scales.

Considering the sky as purely a source of information about the Cosmos,
one could encode this by determining the spectrum in
every direction, or equivalently by measuring
the spectrum for every spherical harmonic coefficient of sky emission
\citep{ScottCNM}.  In 
the Universe (unlike for the Solar System or the Milky Way, or the typical 
experimental laboratory for that matter) there are no special directions -- 
the Universe appears to be isotropic, at least in a statistical sense. 
This means that there are no special $m$s for each multipole $\ell$ (as
there would be in axial symmetry, where $m=0$, for example).  An additionally
important fact is that on sufficiently large scales cosmological structure
is of low contrast and hence 
evolves under linear perturbation theory, which means that there are no 
strong phase correlations and the 2-point correlation function contains the 
majority of the cosmological information. In harmonic space we can therefore 
restrict ourselves to determining the squares of the amplitudes of the
spherical harmonic coefficients summed over $m$ for each
multipole $\ell$:
\begin{equation}
C_{\ell} \equiv \frac{1}{2\ell+1}\sum_{\ell=-m}^{\ell=m} |a_{\ell m}|^{2}.
\end{equation}
It is therefore possible to encompass most of the useful sky information 
by characterizing the frequency spectrum of this fluctuation power spectrum, or 
$C_{\ell}(\nu)$. We show a schematic representation of what these might look 
like in Fig.~\ref{multipolefig}. In a sense this ``power spectrum spectrum'' 
is the blueprint describing just what sort of Universe we live in, and all the 
physical processes operating to give it its large-scale statistical properties. 

Estimating these higher moments is much more challenging than
determining the mean intensity of the sky, owing to the fact that not only
must atmospheric, zodiacal and Galactic contributions be subtracted,
but noise and systematic effects in experimental data must be fully understood
and removed as well.  In general the systematic effects are hard to control
over a wide range of angular scales.
The main exception comes from the CMB, by far the brightest component of the
CB, whose multipole spectrum has been measured to high precision by the
{\it Planck\/} satellite mission \citep{planck2014b}. Further CB 
multipole moments have been estimated in the IR 
\citep{lagache2007,matsuura2011,hajian2012,viero2013,planck2014a}
and the optical 
\citep{kashlinsky2002,thompson2007,cooray2012,zemcov2014,mitchell2015,seo2015}, 
as well as at higher X-ray and $\gamma$-ray energies 
\citep{silwa2001,cappelluti2013,broderick2014}.  However, these are
typically over only a restricted range of scales and there remains much more 
work to be done in order to extract all of the cosmological information
from the background sky.

\begin{figure}[ht!]
\begin{center}
\includegraphics[scale=0.7]{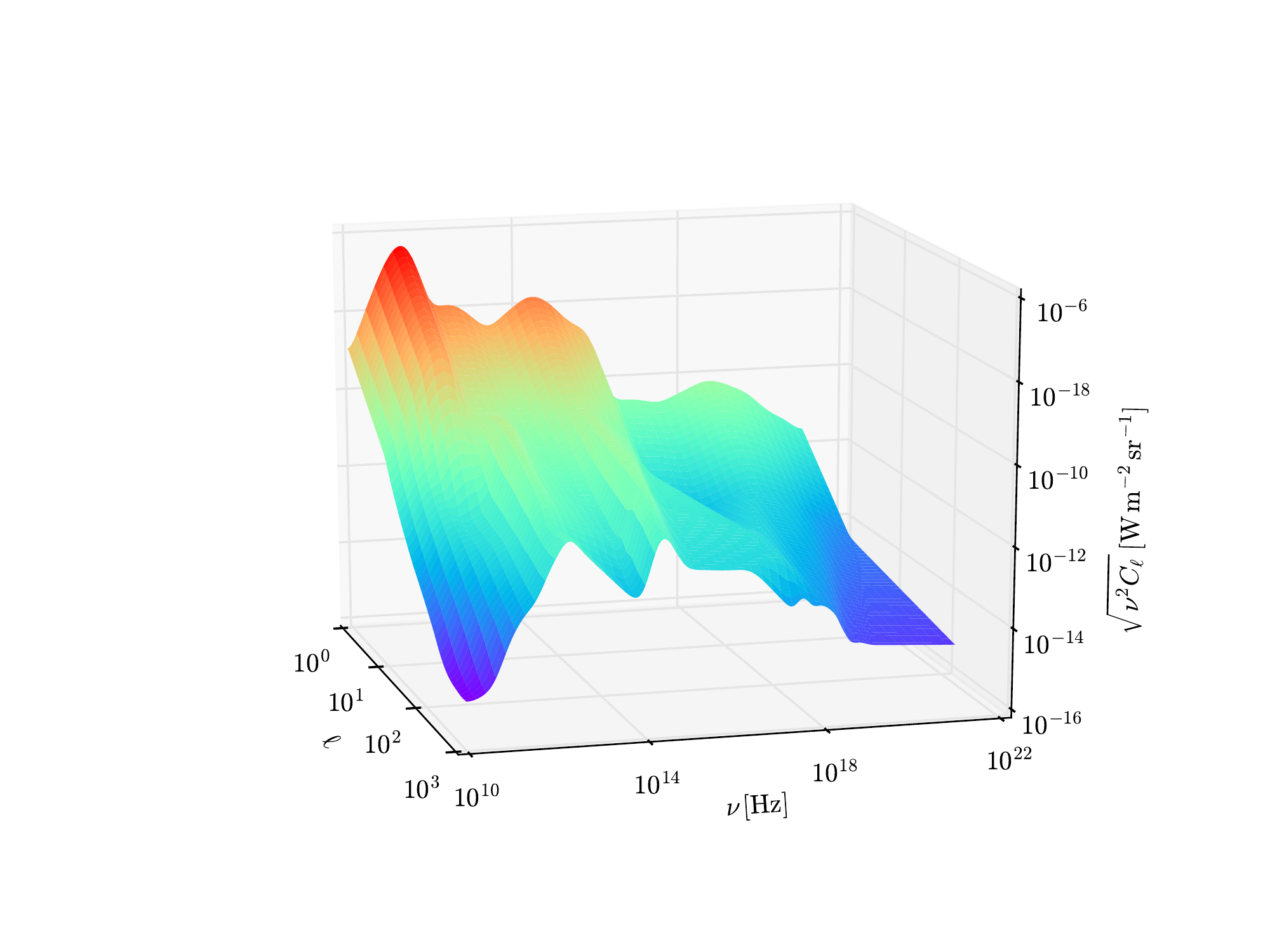}
\caption{\label{multipolefig}
Schematic of the CB as both a function of frequency $\nu$ and
(inverse) angular scale $\ell$. This effectively encodes the entirety of the
useful information in the background sky brightness, telling us all of the
details about the distribution of energy 
and astrophysical processes through the history of our Universe.}
\label{multipolefig}
\end{center}
\end{figure}

\section{Conclusions}
\label{conclusions}

In this paper we have summarized recent measurements of the cosmic
background radiation, in order to 
obtain a comprehensive perspective on the extragalactic photons that 
fill our sky. Spanning 18 orders of magnitude in frequency or wavelength, the 
CB requires a wide range of instruments and detector technologies to measure,
and an understanding of many different astrophysical processes to interpret.

Photons are not the only particles that come to us from the distant
Universe -- there are also cosmic rays, neutrinos and gravitons.  These
can tell us about special places within galaxies (typically neutron stars,
black holes, and galactic nuclei) from where they are emitted.  In future
such non-photon backgrounds might also provide more general information
about cosmology and structure formation.  But for now our view of the
Cosmos is dominated by what we learn from the electromagnetic radiation
background, which has been the subject of our review.

We split the CB into seven components, from radio to $\gamma$-rays, which
we call the CRB, CMB, CIB, COB, CUB, CXB and CGB\@. Very broadly, 
the extragalactic radio emission comes from synchrotron radiation and the 
low-energy tail of the CMB, while the CMB is the redshifted thermal
remnant of the hot early Universe. The CIB comes mainly from dust that
has been heated by stars and re-radiated at cooler temperatures, while 
the COB comes from the stars directly. In the CUB we have a combination of 
scattered starlight and hot diffuse gas, while the CXB comes predominantly
from the accretion disks of galactic nuclei.  Lastly, CGB photons are
primarily emitted from the ultra high-energy jets of galactic nuclei.
In order to determine the 
cosmic average sky brightness, it is necessary to subtract the emission from 
the instrument, the Earth's atmosphere, the Solar System and the Milky Way.
This is challenging 
in most wavebands, and hence our knowledge of the CB is still quite incomplete.

Small amplitude spectral features should be present in the monopole spectrum,
arising from line emission processes, and we have presented order-of-magnitude
estimates for some of the most astrophysically important lines.
The study of such line features is still 
in its infancy, but future experiments hold great promise through 
measurement of structure in 3D data cubes (``intensity mapping'') and of
spatial correlations with other emission processes in the fluctuating sky.

The majority of statistical information present in the CB is encoded in
higher-order multipole moments. 
Precise estimates of the power in these multipoles of the CMB pin down models
of the early Universe, while at other wavebands we have cruder estimates of
the correlated sky that tells us about the formation of structure at
more recent epochs.  One could imagine essentially having a full spectrum
of the power at every angular scale on the sky.  This ``spectrum of the
power spectrum'' would encode a large fraction of the information accessible
to us that characterizes our entire observable Universe.

\section*{Acknowledgements}

This research was supported by the Natural Sciences and Engineering Research Council (NSERC) of Canada.

\bibliography{SOTU_arXiv}

\end{document}